\documentclass[showpacs,preprintnumbers,amsmath,amssymb,12pt,floatfix]{revtex4}
\usepackage[pdftex]{graphicx}
\usepackage[usenames, dvipsnames]{color}
\usepackage{amssymb}
\usepackage{bm}
\usepackage{graphicx}
\usepackage{revsymb}
\usepackage{amsmath}
\usepackage{dcolumn}
\usepackage{ulem}

\usepackage{color} % to paint the words to be indexed

\newcommand{\be}{\begin{eqnarray}}
\newcommand{\ee}{\end{eqnarray}}

\begin{document}
\draft
\title{Gamow-Teller strength distributions of $^{18}$O and well-deformed  nuclei $^{24, 26}$Mg by deformed QRPA}

\author{Eunja Ha \footnote{ejaha@hanyang.ac.kr}}
\address{Department of Physics and Research Institute for Natural Science, Hanyang University, Seoul, 04763, Korea}
\author{Myung-Ki Cheoun \footnote{cheoun@ssu.ac.kr}}
\address{Department of Physics and Origin of Matter and Evolution of Galaxies (OMEG) Institute, Soongsil University, Seoul 156-743, Korea}
\author{H. Sagawa \footnote{sagawa@ribf.riken.jp}}
\address{RIKEN, Nishina Center for Accelerator-Based Science, Wako 351-0198, Japan and Center for Mathematics and Physics, University of Aizu, Aizu-Wakamatsu, Fukushima 965-8560, Japan}
\author{Gianluca Col\`o \footnote{colo@mi.infn.it}}
\address{Dipartimento di Fisica, Universit\`a degli Studi and INFN via Celoria 16, 20133 Milano, Italy}

\begin{abstract}
We investigate the Gamow-Teller (GT) transition strength distributions of {strongly} deformed nuclei, $^{24,26}$Mg, as well as of $^{18}$O.  The calculations are performed within a deformed quasi-particle random phase approximation (DQRPA) which explicitly includes the deformation degree of freedom  in the Skyrme-Hartree-Fock (SHF) and RPA calculations. The residual particle-particle ($p-p$) interaction as well as the particle-hole ($p-h$) interaction are extracted from Br\"uckner $G$-matrix calculations.  The {residual interaction} dependence of the low-lying GT strength of these strongly deformed nuclei is examined by changing the strength of  the residual $p-p$ and  $p-h$ interactions. 
 We have found that the low-lying GT peaks are quite similar in energy to those found in {spherical} $N=Z$ and $N=Z+2$ nuclei near magic shells,  but the configurations {of $^{24,26}$Mg are largely mixed by} the pairing correlations and the deformation.   Our results are compared to the experimental GT $(\pm)$ transition data by ($t$, $^3$He) and ($^{3}$He, $t$) reactions, {and found to reproduce the main features of GT strength distributions.}
\end{abstract}

\date{\today}

\maketitle

\section{Introduction}

The Gamow-Teller (GT) excitation is one of the key transitions when  studying the weak interaction of nuclei through charge-exchange (CEX) (or charged-current (CC)) reactions, beta decays, and electron or muon capture reactions. Recent remarkable progresses in the study of supernova (SN) neutrinos reveal that the GT transition is a key transition in the neutrino-induced reactions relevant to the neutrino processes   \cite{Taka2015,Bala2015,Taka2013,Cheoun2012,Moha2023,Suzuki2023,Cheoun2023}. For example, the GT transition turns out to be dominant among various multipole transitions in the neutrino-nucleus ($\nu$-A) reaction, according to recent theoretical calculations \cite{Taka2015,Bala2015}. However, we do not have enough data to confirm the reliability of  theoretical models. Only limited data for the neutrino-induced reactions are available until now, despite many discussions about the possibility of low-energy neutrino sources \cite{Sato2005,Shin2016}. The recent progress in JSNS$^2$ experiments \cite{JSNS2} could provide valuable low-energy neutrino beams on nuclear targets.

In this respect, neutrino reaction data at LSND \cite{LSND1,LSND2}, and recent successful CEX reaction data by proton or neutron beam at RIKEN or NSCL \cite{Sasano2011,Sasano2012}, triton and {$^{3}$He} beam at RCNP \cite{Freke2016}, and other light nuclei beams at other facilities, are greatly helpful for understanding the neutrino-induced reactions in the following aspects.
First, the main contribution to the neutrino-induced reaction, namely, the GT transition, can be studied experimentally by the CEX reaction. Second, the quenching factor can be deduced from the GT 
data. {Here, we should note that recent {{\it ab initio}} calculations  have explained the quenching factor {in a number of light- and
medium-mass nuclei by the combined effect of two-body currents as well as by} 
strong correlations \cite{Gysb2019,Heiko2020}.}  Third, some high-lying GT excitations, which  contribute to the neutrino-induced reactions, can be investigated by the CEX reactions. These CEX data may also provide important information for the $r$-process nucleosynthesis study, because additional protons and/or neutrons in cosmological nucleosynthesis sites may change the nuclear species by means of CEX reactions.
For example, Gamow-Teller (+) transitions from $^{24}$Mg affect the electron capture rates in the O+Ne+Mg cores of stars, but theoretical results in Refs. \cite{Full1982,Jame2007} show some inconsistency {among them}. The understanding of decay branching ratios of the excited $^{24}$Mg states are {also} important for $^{12}$C + $^{12}$C  fusion reactions {\cite{Muns2017,Tumi2018,Monp2022}}. The GT results for $^{18}$O are an important %for an important %nuclear
 input  for astrophysical calculations such as the CNO cycle \cite{Moha2023}, and also for neutrino detection due to the admixture of $^{18}$O in natural water at Hyper-Kamiokande  \cite{Abe2021}.

From the nuclear physics viewpoint, the GT transition is an allowed spin-isospin excitation of a nucleus, and the transition operator is one of the simplest. Consequently, it may give invaluable information of the nuclear spin-isospin structure. However, there still remain some ambiguities, that prevent us from having a full quantitative understanding of GT excitations throughout the nuclear chart; the same issues prevent us from knowing precisely the 
%which are also closely related to the ambiguities intensively discussed in the 
matrix elements for {2$\nu\beta\beta$} and  {0$\nu\beta\beta$} decays. 
%They need a more careful approach and deliberate identification of experimental data as well as theoretical calculations. 
In this work, we focus on one of the ambiguities, or open questions, {regarding the} {GT strength in deformed nuclei:} we study the deformation effects on the GT strength, {by using} Deformed Quasi-particle Random Phase Approximation (DQRPA) calculations. 

Many nuclei in the nuclear chart are considered to be  deformed, which can be verified from the E2 transition data as well as the rotational band structure. In particular, neutron-rich and neutron-deficient nuclei may be largely deformed, and these nuclei are among those that have strong low-energy GT strength or short $\beta$-decay half-lives.  Moreover, the deformation may produce unusual phenomena of shell evolution of protons and neutrons {\cite{Ichik2019}}, which make new sub-magic numbers and change to some extent the nature of  particle-particle ($p-p$) and  particle-hole ($p-h$) interactions. One of the practical but successful ways for properly describing the deformation is to start from the Nilsson model with the axial symmetry. Our previous calculations \cite{Ha2015a,Ha2015b,Ha2016} for the GT strength distribution of open shell nuclei by deformed QRPA showed that the deformation markedly alters the GT strength peaks obtained by a spherical QRPA. 

%{Comment: there is a long series of papers by P. Sarriguren and co-workers, that deal with GT distributions and beta-decay of deformed nuclei, and some should be quoted. More recently, K. Yoshida worked on 
%deformed RPA, and perhaps I am forgetting 1-2 more groups. I can help to put these references in due course.}

The $pn$-QRPA  model has  been exploited using  the Skyrme Hartree-Fock (SHF) plus BCS scheme  developed by P. Sarrgiuren and co-workers \cite{Sarri2001,Sarri2001-b,Sarri2005,Sarri2014}. They investigated the GT strength distribution  in Zr and Mo isotopes \cite{Sarri2014}, the neutron-deficient Pb isotopes \cite{Sarri2005}, and the proton-rich Kr isotopes \cite{Sarri2001-b}. For the neutron-rich nuclei, K. Yoshida also employed a model with Skyrme-type energy density functional (EDF) in the DQRPA \cite{Yoshi2008,Yoshi2013}. 

The importance of the residual interaction was discussed in our early reports through calculations of both single- and double-$\beta$ decay transitions \cite {Ch93,pan}. It should be noticed that these calculations were performed using spherical QRPA, which did not include the deformation explicitly.
Primary aim of the present work is to perform the deformed QRPA (DQRPA) calculations with a realistic residual interaction, {determined from  the Br\"uckner $G$-matrix with} the CD Bonn potential. {We utilized as a starting point the Skyrme-type mean field (MF) \cite{Stoitsov} instead of the Woods-Saxon potential employed in the previous calculations \cite{Ha2015a,Ha2015b,Ha2016}.} Therefore, this work is an extension of our recent works \cite{Ha2015a,Ha2015b} for the DQRPA, in which all effects of the deformation are consistently treated in the QRPA framework based on the Woods-Saxon mean field \cite{Ha2015a}. 

As an application of the present DQRPA model, we choose the GT transitions of two Mg isotopes, $^{24}$Mg and $^{26}$Mg, because they are known as well-deformed nuclei. Moreover, there are  precise GT experimental data by $^3$He and $t$ beams in these nuclei \cite {Zegers08,Madey87,Zegers06}. {We perform also the study of an almost spherical nucleus $^{18}$O in DHF calculation as a counter example to strongly deformed nuclei. Our paper is organized as follows.}
In Sec. II, we briefly explain the formalism including the deformation. The applications to the GT transition strength distributions for Mg isotopes are performed in Sec. III with  detailed discussions on the choice of  necessary parameters for $p-p$ and   $p-h$  residual interactions in the present formalism. Conclusions are drawn with a short {perspective for} future works in Sec. IV.

\section{formalism}

Our calculation is carried out by using the QRPA \cite{Ha2015a}, which adopts the SHF mean field \cite{Stoitsov} for the mean field, and the residual interactions calculated by the Br\"uckner $G$-matrix based on the CD Bonn potential.  On top of the mean field, the pairing correlations are taken into account in the BCS approximation.  We refer to our previous papers for the detailed calculations of the BCS wave functions including the tensor force (TF) \cite{Ha18-1,Ha18-2}. Hereafter, we briefly summarize the DQRPA model, which will be applied for calculations of the GT strength distributions.

In the following, we adopt the standard QRPA formalism based on the equation of motion for the following phonon operator, acting on the BCS ground state \cite{Ha2015a}: 
\begin{equation}\label{phonon}
{\cal Q}^{\dagger}_{m,K}  =\sum_{\rho_{\alpha} \alpha \alpha'' \rho_{\beta} \beta \beta''}
[ X^{m}_{( \alpha \alpha'' \beta \beta'')K} A^{\dagger}( \alpha \alpha'' \beta \beta'' K)
- Y^{m}_{( \alpha \alpha'' \beta \beta'')K} {\tilde A}( \alpha \alpha'' \beta \beta'' K)]~.
\end{equation}
{where $\rho_{\alpha (\beta)} (= \pm 1)$ denotes the sign of the total angular momentum projection of the $\alpha$ state for the reflection symmetry.}
Here, we have introduced pair creation and annihilation operators, composed by two quasiparticles and defined as
\begin{equation}
 A^{\dagger}( \alpha \alpha'' \beta \beta'' K)  =
 {[a^{\dagger}_{ \alpha \alpha''} a^{\dagger}_{\beta \beta''}]}^K,
~~~{\tilde A}( \alpha \alpha'' \beta \beta'' K)  =
 {[a_{\beta \beta''} a_{\alpha \alpha''}]}^K,
\end{equation}
where $K$ is the quantum number associated with the projection of the intrinsic angular momentum on the symmetry axis, which is a good quantum number in the axially deformed nuclei.
We note that parity is also treated as a good quantum number in the present approach.  
Here, $\alpha$ indicates a set of quantum numbers to specify the single-particle-state (SPS). Isospin of the real particle is denoted by the Greek letter with prime $(\alpha' , \beta' , \gamma' , \delta')$ (see Eqs. (\ref{eq:mat_A}) and (\ref{eq:mat_B})). {Since our formalism is constructed to include the $np$ pairing} correlations composed of $T=0$ and $T=1$ contributions, we have two different types of quasiparticles, quasi-proton and quasi-neutron, and the isospin of the quasiparticles cannot be clearly defined \cite{Ha2015a}. The {quantum number corresponding to the quasi-isospin is denoted by $(\alpha'', \beta'', \gamma'', \delta'')$, that is, with double-primes: in other words, these indices are %we denote the quasiparticles as 
$\alpha'' ( \beta'', \gamma'', \delta'') = 1,2$ and are used hereafter, instead of $\tau_z=\pm1$ for neutrons and protons.} Detailed discussion regarding the $T=0$ part has been done in our previous papers \cite{Ha18-1,Ha18-2}. 

Within the quasi-boson approximation for the phonon operator, we obtain the following QRPA equation for describing the correlated DQRPA ground state:

\begin{eqnarray}\label{qrpaeq}
&&\left(\begin{array}{cccccccc}
           A_{\alpha \beta \gamma \delta (K)}^{1111} & A_{\alpha \beta \gamma \delta (K)}^{1122} &
           A_{\alpha \beta \gamma \delta (K)}^{1112} & A_{\alpha \beta \gamma \delta (K)}^{1121} &
           B_{\alpha \beta \gamma \delta (K)}^{1111} & B_{\alpha \beta \gamma \delta (K)}^{1122} &
           B_{\alpha \beta \gamma \delta (K)}^{1112} & B_{\alpha \beta \gamma \delta (K)}^{1121} \\
           A_{\alpha \beta \gamma \delta (K)}^{2211} & A_{\alpha \beta \gamma \delta (K)}^{2222} &
           A_{\alpha \beta \gamma \delta (K)}^{2212} & A_{\alpha \beta \gamma \delta (K)}^{2221} &
           B_{\alpha \beta \gamma \delta (K)}^{2211} & B_{\alpha \beta \gamma \delta (K)}^{2222} &
           B_{\alpha \beta \gamma \delta (K)}^{2212} & B_{\alpha \beta \gamma \delta (K)}^{2221}\\
           A_{\alpha \beta \gamma \delta (K)}^{1211} & A_{\alpha \beta \gamma \delta (K)}^{1222} &
           A_{\alpha \beta \gamma \delta (K)}^{1212} & A_{\alpha \beta \gamma \delta (K)}^{1221} &
           B_{\alpha \beta \gamma \delta (K)}^{1211} & B_{\alpha \beta \gamma \delta (K)}^{1222} &
           B_{\alpha \beta \gamma \delta (K)}^{1212} & B_{\alpha \beta \gamma \delta (K)}^{1221}\\
           A_{\alpha \beta \gamma \delta (K)}^{2111} & A_{\alpha \beta \gamma \delta (K)}^{2122} &
           A_{\alpha \beta \gamma \delta (K)}^{2112} & A_{\alpha \beta \gamma \delta (K)}^{2121} &
           B_{\alpha \beta \gamma \delta (K)}^{2111} & B_{\alpha \beta \gamma \delta (K)}^{2122} &
           B_{\alpha \beta \gamma \delta (K)}^{2112} & B_{\alpha \beta \gamma \delta (K)}^{2121} \\
             &       &       &      &      &        &           &        \\ \nonumber
           - B_{\alpha \beta \gamma \delta (K)}^{1111} & -B_{\alpha \beta \gamma \delta (K)}^{1122} &
            -B_{\alpha \beta \gamma \delta (K)}^{1112} & -B_{\alpha \beta \gamma \delta (K)}^{1121} &
           - A_{\alpha \beta \gamma \delta (K)}^{1111} & -A_{\alpha \beta \gamma \delta (K)}^{1122} &
           -A_{\alpha \beta \gamma \delta (K)}^{1112}  & -A_{\alpha \beta \gamma \delta (K)}^{1121}\\
           - B_{\alpha \beta \gamma \delta (K)}^{2211} & -B_{\alpha \beta \gamma \delta (K)}^{2222} &
           -B_{\alpha \beta \gamma \delta (K)}^{2212}  & -B_{\alpha \beta \gamma \delta (K)}^{2221} &
           - A_{\alpha \beta \gamma \delta (K)}^{2211} & -A_{\alpha \beta \gamma \delta (K)}^{2222} &
           -A_{\alpha \beta \gamma \delta (K)}^{2212}  & -A_{\alpha \beta \gamma \delta (K)}^{2221}\\
           - B_{\alpha \beta \gamma \delta (K)}^{1211} & -B_{\alpha \beta \gamma \delta (K)}^{1222} &
           -B_{\alpha \beta \gamma \delta (K)}^{1212}  & -B_{\alpha \beta \gamma \delta (K)}^{1221} &
           - A_{\alpha \beta \gamma \delta (K)}^{1211} & -A_{\alpha \beta \gamma \delta (K)}^{1222} &
           -A_{\alpha \beta \gamma \delta (K)}^{1212}  & -A_{\alpha \beta \gamma \delta (K)}^{1221} \\
          - B_{\alpha \beta \gamma \delta (K)}^{2111} & -B_{\alpha \beta \gamma \delta (K)}^{2122} &
           -B_{\alpha \beta \gamma \delta (K)}^{2112}  & -B_{\alpha \beta \gamma \delta (K)}^{2121} &
           - A_{\alpha \beta \gamma \delta (K)}^{2111} & -A_{\alpha \beta \gamma \delta (K)}^{2122} &
           -A_{\alpha \beta \gamma \delta (K)}^{2112}  & -A_{\alpha \beta \gamma \delta (K)}^{2121} \\
           \end{array} \right)\\  && \times
\left( \begin{array}{c}   {\tilde X}_{(\gamma 1 \delta 1)K}^{m}  \\ {\tilde X}_{(\gamma 2 \delta 2)K}^{m} \\
  {\tilde X}_{(\gamma 1 \delta 2)K}^{m} \\  {\tilde X}_{(\gamma 2 \delta 1)K}^{m} \\ \\
     {\tilde Y}_{(\gamma 1 \delta 1)K}^{m} \\ {\tilde Y}_{(\gamma 2 \delta 2)K}^{m} \\
     {\tilde Y}_{(\gamma 1 \delta 2)K}^{m}\\{\tilde Y}_{(\gamma 2 \delta 1)K}^{m} \end{array} \right)
 = \hbar {\Omega}_K^{m}
 \left ( \begin{array}{c} {\tilde X}_{(\alpha 1 \beta 1)K}^{m}  \\{\tilde X}_{(\alpha 2 \beta 2)K}^{m} \\
 {\tilde X}_{(\alpha 1 \beta 2)K}^{m} \\  {\tilde X}_{(\alpha 2 \beta 1)K}^{m}\\ \\
{\tilde Y}_{(\alpha 1 \beta 1)K}^{m} \\ {\tilde Y}_{(\alpha 2 \beta 2)K}^{m} \\
{\tilde Y}_{(\alpha 1 \beta 2)K}^{m} \\ {\tilde Y}_{(\alpha 2 \beta 1)K}^{m} \end{array} \right)  ~,
\end{eqnarray}
where the amplitudes
${\tilde X}^m_{(\alpha \alpha''  \beta \beta'')K }$ and ${\tilde Y}^m_{(\alpha
\alpha''  \beta \beta'')K}$ in Eq. (\ref{qrpaeq}) stand for forward and backward going amplitudes from the state ${ \alpha
\alpha'' }$ to the state  ${\beta  \beta''}$ \cite{Ha2015a}, 

The $A$ and $B$ matrices in Eq. (\ref{qrpaeq}) are given by
\begin{eqnarray} \label{eq:mat_A}
A_{\alpha \beta \gamma \delta (K)}^{\alpha'' \beta'' \gamma'' \delta''}  = && (E_{\alpha
   \alpha''} + E_{\beta \beta''}) \delta_{\alpha \gamma} \delta_{\alpha'' \gamma''}
   \delta_{\beta \delta} \delta_{\beta'' \delta''}
       - \sigma_{\alpha \alpha'' \beta \beta''}\sigma_{\gamma \gamma'' \delta \delta''}\\ \nonumber
   &\times&
   \sum_{\alpha' \beta' \gamma' \delta'}
   [-g_{pp} (u_{\alpha \alpha''\alpha'} u_{\beta \beta''\beta'} u_{\gamma \gamma''\gamma'} u_{\delta \delta''\delta'}
   +v_{\alpha \alpha''\alpha'} v_{\beta \beta''\beta'} v_{\gamma \gamma''\gamma'} v_{\delta \delta''\delta'} )
    ~V_{\alpha \alpha' \beta \beta',~\gamma \gamma' \delta \delta'}
    \\ \nonumber  &-& g_{ph} (u_{\alpha \alpha''\alpha'} v_{\beta \beta''\beta'}u_{\gamma \gamma''\gamma'}
     v_{\delta \delta''\delta'}
    +v_{\alpha \alpha''\alpha'} u_{\beta \beta''\beta'}v_{\gamma \gamma''\gamma'} u_{\delta \delta''\delta'})
    ~V_{\alpha \alpha' \delta \delta',~\gamma \gamma' \beta \beta'}
     \\ \nonumber  &-& g_{ph} (u_{\alpha \alpha''\alpha'} v_{\beta \beta''\beta'}v_{\gamma \gamma''\gamma'}
     u_{\delta \delta''\delta'}
     +v_{\alpha \alpha''\alpha'} u_{\beta \beta''\beta'}u_{\gamma \gamma''\gamma'} v_{\delta \delta''\delta'})
    ~V_{\alpha \alpha' \gamma \gamma',~\delta \delta' \beta \beta' }],
\end{eqnarray}
\begin{eqnarray} \label{eq:mat_B}
B_{\alpha \beta \gamma \delta (K)}^{\alpha'' \beta'' \gamma'' \delta''}  =
 &-& \sigma_{\alpha \alpha'' \beta \beta''} \sigma_{\gamma \gamma'' \delta \delta''}
  \\ \nonumber &\times&
 \sum_{\alpha' \beta' \gamma' \delta'}
  [g_{pp}
  (u_{\alpha \alpha''\alpha'} u_{\beta \beta''\beta'}v_{\gamma \gamma''\gamma'} v_{\delta \delta''\delta'}
   +v_{\alpha \alpha''\alpha'} v_{{\bar\beta} \beta''\beta'}u_{\gamma \gamma''\gamma'} u_{{\bar\delta} \delta''\delta'} )
   ~ V_{\alpha \alpha' \beta \beta',~\gamma \gamma' \delta \delta'}\\ \nonumber
     &- & g_{ph} (u_{\alpha \alpha''\alpha'} v_{\beta \beta''\beta'}v_{\gamma \gamma''\gamma'}
     u_{\delta \delta''\delta'}
    +v_{\alpha \alpha''\alpha'} u_{\beta \beta''\beta'}u_{\gamma \gamma''\gamma'} v_{\delta \delta''\delta'})
   ~ V_{\alpha \alpha' \delta \delta',~\gamma \gamma' \beta \beta'}
     \\ \nonumber  &- & g_{ph} (u_{\alpha \alpha''\alpha'} v_{\beta \beta''\beta'}u_{\gamma \gamma''\gamma'}
      v_{\delta \delta''\delta'}
     +v_{\alpha \alpha''\alpha'} u_{\beta \beta''\beta'}v_{\gamma \gamma''\gamma'} u_{\delta \delta''\delta'})
   ~ V_{\alpha \alpha' \gamma \gamma',~\delta \delta' \beta \beta'}],
\end{eqnarray}
where $\sigma_{\alpha \alpha'' \beta \beta''}$ = 1 if $\alpha = \beta$ and $\alpha''$ =
$\beta''$, otherwise $\sigma_{\alpha \alpha'' \beta \beta'' }$ = $\sqrt 2$ \cite{Ch93}.  The $u$ and $v$ coefficients are determined from the gap equations. The $g_{pp}$ and $g_{ph}$ stand for particle-particle and particle-hole renormalization factors for the residual interactions in Eqs. (\ref{eq:mat_A}) and (\ref{eq:mat_B}).  The two-body interactions $V_{\alpha \beta,~\gamma \delta}$ and $V_{\alpha \delta,~\gamma \beta}$ are particle-particle and particle-hole matrix elements of the residual $N$-$N$ interaction $V$, respectively, which are calculated from the $G$-matrix {as solutions of the Bethe-Goldstone equation from the CD Bonn potential.}  

{The two-body interactions $V_{\alpha \beta,~\gamma \delta}$  and $V_{\alpha \delta,~\gamma \beta}$ are, respectively, $p-p$ and $p-h$ matrix elements of the residual $N-N$ interaction in the deformed state. They are 
calculated from the $G$-matrix in the spherical basis as follows
\begin{eqnarray}
V_{\alpha \alpha' \beta \beta' ,~\gamma \gamma' \delta \delta'} 
= - && \sum_{J} \sum_{abcd} F_{\alpha a {\bar\beta} b}^{JK}  F_{\gamma c {\bar \delta} d}^{JK} G( a \alpha' b \beta' c \gamma' d \delta' , J)~, \\ \nonumber
V_{\alpha \alpha' \delta \delta' , \gamma \gamma' \beta \beta'} 
= && \sum_J \sum_{abcd} F_{\alpha a {\delta} d}^{JK}  F_{\gamma
 c {\beta} b}^{JK} G( a \alpha' d \delta' c \gamma' b \beta' , J)~, \\ \nonumber
V_{\alpha \alpha' \gamma \gamma' , \delta \delta' \beta \beta'} 
 = && \sum_J \sum_{abcd} F_{\alpha a {\gamma} c}^{JK}  F_{\beta
 b {\delta} d}^{JK} G( a \alpha' c \gamma' d \delta' b \beta' , J)~.
\end{eqnarray} 
Here $a$ and $\alpha$ indicates spherical and deformed SPS, respectively. {We note that the quasi-isospin $\alpha', \beta', \gamma'$ and $\delta'$ in the $G$-matrix is {replaced} by total isospin ($T=0$ or $T=1$) of the two-body interaction in the isospin representation.} We use $F_{\alpha a {\bar \beta} b}^{JK} = B_{a}^{\alpha} B_b^{\beta} C_{j_a \Omega_\alpha j_b \Omega_\beta}$ for $K = \Omega_{\alpha} + \Omega_{\beta}$. The expansion coefficient $B_{\alpha}^a$ is defined as 
\begin{equation}
| \alpha \Omega_{\alpha} > =\sum_{a} B_{a}^{\alpha} |a \Omega_{\alpha} > ~,~  B_{a}^{\alpha} = \sum_{N n_z} C_{l \Lambda { 1 \over 2} \Sigma}^{j \Omega_\alpha} A_{N n_z \Lambda}^{ N_0 l} b_{N n_z \Sigma},  
\end{equation}
with the Clebsch-Gordan coefficient $C_{l \Lambda { 1 \over 2} \Sigma}^{j \Omega_\alpha}$, the spatial overlap integral $A_{N n_z \Lambda}^{ N_0 l}$, and the eigenvalues obtained from the total Hamiltonian in the deformed basis $b_{N n_z \Sigma}$. Detailed formulas regarding the transformation of Eq.~(7) can be found in Ref. \cite{Ha2015a}.}

Our DQRPA equation has a very general form because we include the deformation as well as two  kinds of pairing correlations, $(T=1, S=0)$ and $(T=0, S=1)$, in the model. 
If we switch off the $np$ pairing, all off-diagonal terms in the $A$ and $B$ matrices in Eq. (\ref{qrpaeq}) disappear with the replacement of 1 and 2 into $p$ and $n$. Then the DQRPA equation is decoupled into $pp + nn + pn + np$ DQRPA equations. $pp + nn$ QRPA can describe charge-conserving reactions such as the electromagnetic (EM) transitions, including magnetic dipole (M1) transition, between the initial and final states in  the same nucleus, while $np + pn$ QRPA describes charge-exchange reactions like  the GT${(+/-)}$  transitions between mother and daughter nuclei having different proton and neutron numbers, $(N,Z)\rightarrow (N\pm1, Z\mp1)$. 
If we assume spherical symmetry, our equations are reduced to 
%Namely, if we use the spherical QRPA, this equation is reduced to 
the QRPA equations in Ref. \cite{Ch93}. If we neglect the $np$ pairing for {the nuclei considered in this work,} {\it i.e.} take only 1212 terms, Eq. (\ref{qrpaeq}) becomes the proton-neutron DQRPA 
as in Ref. \cite{saleh}. %, and quasi-particle 1 and 2 reduce to quasi-proton and quasi-neutron.

The GT transition operator ${\hat {\textrm{O}} }_{1\mu}^{\pm}$ is defined by
 \begin{equation} \label{eq:btop}
{\hat {\textrm{O}} }_{1 \mu}^{-}  = \sum_{\alpha \beta }
< \alpha p |\tau^{+} \sigma_K | \beta n >  c_{\alpha p}^{\dagger} {\tilde c}_{\beta n},~
{\hat {\textrm{O}}}_{1 \mu}^{+}  =  {( { \hat T}_{1 \mu}^{-} )}^\dagger  = {(-)}^{\mu}
{ \hat {\textrm{O}}}_{1,-\mu}^{-}.
\end{equation}
Detailed calculation for the transformation from the intrinsic frame to the nuclear laboratory system were presented in Ref. \cite{Ha2015a}. 
The ${\hat {\textrm{O}}}^{\pm}$ transition amplitudes from the ground state of an initial (parent) nucleus
to the excited state of a final (daughter) nucleus, {\it i.e.} the one phonon state
$\vert K^+, m\rangle$, are written as
\begin{eqnarray} \label{eq:phonon}
&&< K^+, m | {\hat {\textrm{O}}}_{K }^- | ~QRPA >  \\ \nonumber
&&= \sum_{\alpha \alpha''\rho_{\alpha} \beta \beta''\rho_{\beta}}{\cal N}_{\alpha \alpha''\rho_{\alpha}
 \beta \beta''\rho_{\beta} }
 < \alpha \alpha''p \rho_{\alpha}|  \sigma_K | \beta \beta''n \rho_{\beta}>
 [ u_{\alpha \alpha'' p} v_{\beta \beta'' n} X_{(\alpha \alpha''\beta \beta'')K}^{m} +
v_{\alpha \alpha'' p} u_{\beta \beta'' n} Y_{(\alpha \alpha'' \beta \beta'')K}^{m}], \\ \nonumber
&&< K^+, m | {\hat {\textrm{O}}}_{K }^+ | ~QRPA >  \\ \nonumber
&&= \sum_{\alpha \alpha'' \rho_{\alpha} \beta \beta''\rho_{\beta}}{\cal N}_{\alpha \alpha'' \beta \beta'' }
 < \alpha \alpha''p \rho_{\alpha}|  \sigma_K | \beta \beta''n \rho_{\beta}>
 [ u_{\alpha \alpha'' p} v_{\beta \beta'' n} Y_{(\alpha \alpha'' \beta \beta'')K}^{m} +
v_{\alpha \alpha'' p} u_{\beta \beta'' n} X_{(\alpha \alpha'' \beta \beta'')K}^{m} ]~,
\end{eqnarray}
where $|~QRPA >$ denotes the correlated QRPA ground state in the intrinsic frame and
the nomalization factor is given as $ {\cal N}_{\alpha \alpha'' \beta
 \beta''} (J) = \sqrt{ 1 - \delta_{\alpha \beta} \delta_{\alpha'' \beta''} (-1)^{J + T} }/
 (1 + \delta_{\alpha \beta} \delta_{\alpha'' \beta''}).$ 

Here, we use the following %energy density functional for the 
pairing interaction, 
\be
{\cal H}_{pair} ({\bf r}) = {1 \over 2} V_0 [ 1 - V_1 ({\rho / \rho_0  })^{\gamma} ], 
\ee
where the parameter $V_0$ for $^{18}$O is adjusted for fixing the GT peak position, and the other $V_0$, $V_1$ and $\gamma$ are taken from Ref. \cite{Stoitsov}. The results of the parameters and the pairing gaps  obtained with the pairing window {associated with an energy} cut-off of 60 MeV, are summarized in Table \ref{tab:pairing} for {the Skyrme EDF parameter sets adopted in the present work.}   
\begin{table}
\caption[bb]{The parameters relevant to the pairing interaction in the present calculation together with the pairing gaps. \\}
\setlength{\tabcolsep}{2.0 mm}
\begin{tabular}{ccccccc}\hline
 Nucleus & Skyrme  EDF & $V_0$ & {$V_1$} & {$\gamma$} &$\Delta_p$ [MeV] & $\Delta_n$ [MeV]  \\ \hline \hline
$^{18}$O & SLy4    & -125 & 0.5& 1.0 &0.03 & 0.13    \\
$^{24}$Mg & SkP    & -213 & 0.5& 1.0& 0.178 & 0.184    \\
$^{26}$Mg & SkP     & -213 & 0.5& 1.0&0.149 & 0.300   \\ \hline
 \end{tabular}
\label{tab:pairing}
\end{table}

\begin{figure}
\includegraphics[width=0.65\linewidth]{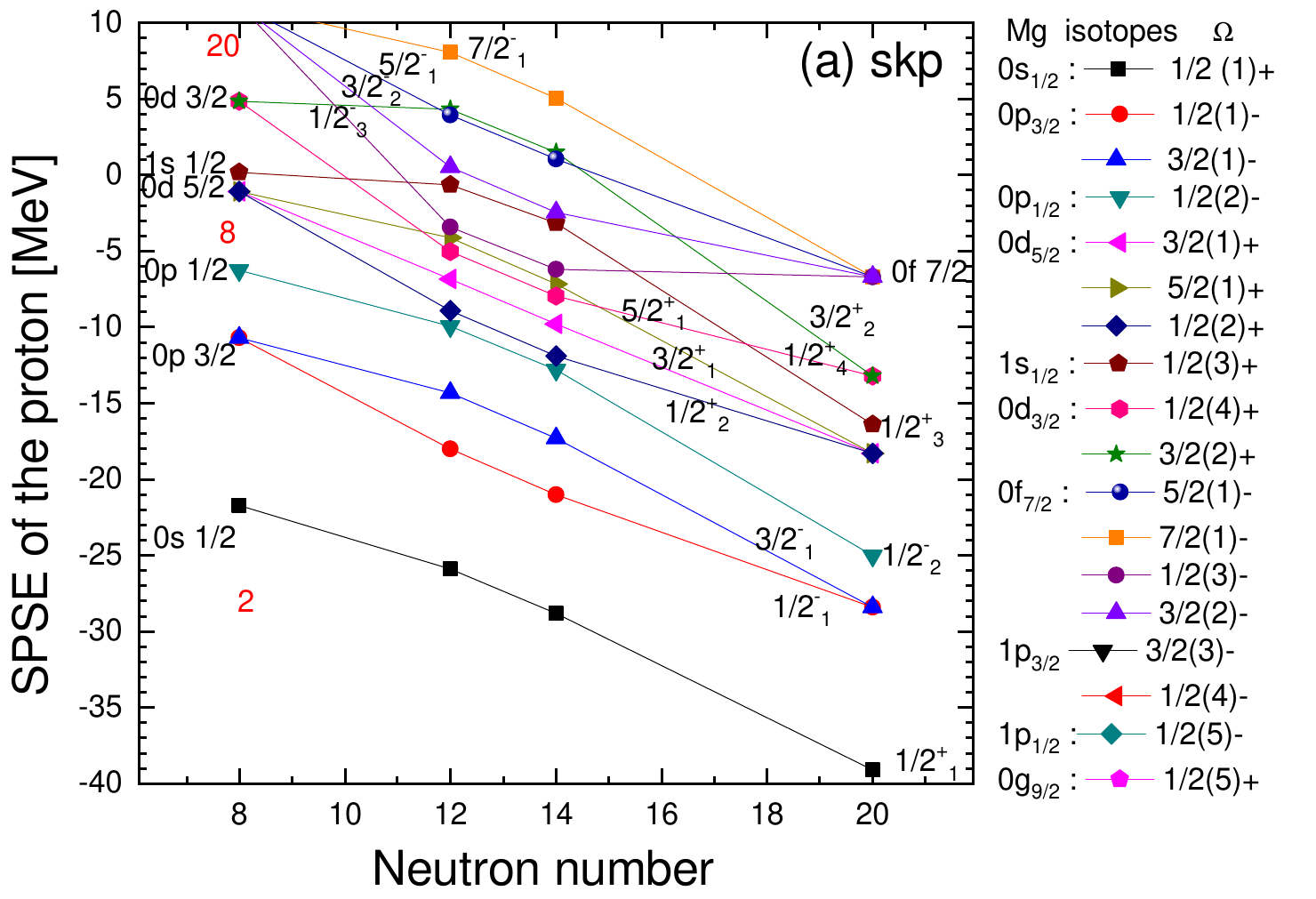}
\includegraphics[width=0.65\linewidth]{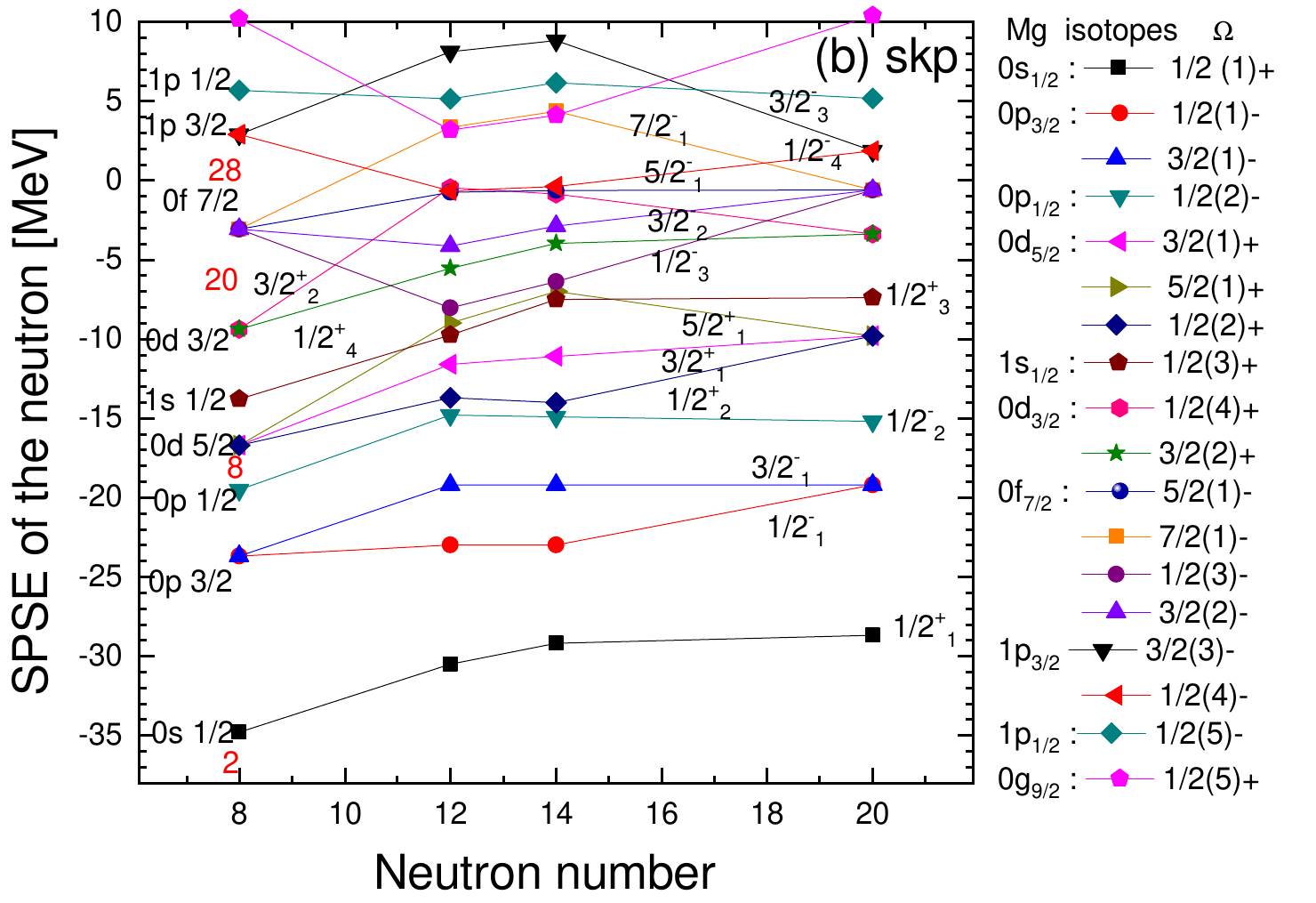}
\caption{(Color online)  Evolution of proton (a) and neutron (b) SPS energies in Mg isotopes by using the SkP parameter set in the SHFB approach. 
The deformations determined by the minimal ground state energy are changed as $\beta_2= 0.03,  0.45, 0.45, 0.00$, respectively, for $N= 8, 12, 14, 20$.}
\label{fig:SPS}
\end{figure}
\section{Results and discussions}
This study is based on the mean field obtained with the axially deformed Skyrme Hartree-Fock-Bogoliubov (SHFB) approach using a harmonic oscillator basis \cite{Stoitsov}. The particle model space, for the nuclei considered here, {includes states} up to $N = 5 \hbar \omega$ and $N = 10 \hbar \omega$, respectively, for the deformed and spherical basis. In the DQRPA, we adopt the wave functions and the SP energies from the SHFB equations.  
We utilize the SP energies from the deformed SHF mean field, and the $G$-matrix for the two-body interaction in Eq. (\ref{eq:mat_A}) and (\ref{eq:mat_B}) is also calculated by using the wave functions  from the deformed SHF.

In Fig.~\ref{fig:SPS}, we illustrate the SPS evolution of protons and neutrons with the deformation. We note that the SP energy splitting of either the $d_{3/2}$ state or the $d_{5/2}$ state, 
induced by deformation in $^{24,26}$Mg, smears out the shell closures, and the magic numbers $N=8$ and 20 are disappeared. 
 Instead,  in the current work, $N = 12$ appears as a proton {quasi-magic number} for $^{24}_{12}$Mg$_{12}$ and $^{26}_{12}$Mg$_{14}$ because of  the relatively large energy gaps above the $N=12$ shell induced by the strong deformations as tabulated in Table \ref{tab:beta2}. {We note that the pairing interaction in Eq. (10) may smear the occupation probabilities, but the smearing is not significant, as shown in Fig. \ref{fig:26mg_OP}.}

Another interesting point is that the SP energies of $\nu d_{5/2}$ and $\nu d_{3/2}$ {orbitals} are slightly higher in energy as the neutron number is increased while  those of protons $\pi d_{5/2}$ and $\pi d_{3/2}$ orbitals are decreased. As a consequence, the Fermi energy gap of protons and neutrons for $^{26}$Mg and $^{16}$O is about 6 MeV, but for $^{24}$Mg it is about 1.5 MeV. Interestingly, in $^{24,26}$Mg, the three states,  $\nu ({5/2}^+_1)$ and $\nu ({1/2}^+_3)$ as well as $\nu ({1/2}^-_3)$, are almost degenerate. 

\subsection{$^{24}$Mg}
\begin{table}
\caption[bb]{Deformation parameters at the ground state minimal energy and Fermi energies for the three nuclei.}
\setlength{\tabcolsep}{2.0 mm}
\begin{tabular}{cccc}\hline
 Nucleus & $\beta_2^{Ours}$ & $\epsilon_n$ (MeV)& $\epsilon_p$ (MeV) \\ \hline \hline
$^{18}$O & 0.03    & - 6.51 & - 11.31    \\
$^{24}$Mg & 0.45    & { - 10.81} &  {- 6.06} \\
$^{26}$Mg & 0.45    &  {- 7.49 } &{- 8.97}   \\ \hline
 \end{tabular}
\label{tab:beta2}
\end{table}

\begin{figure}
\includegraphics[width=0.45\linewidth]{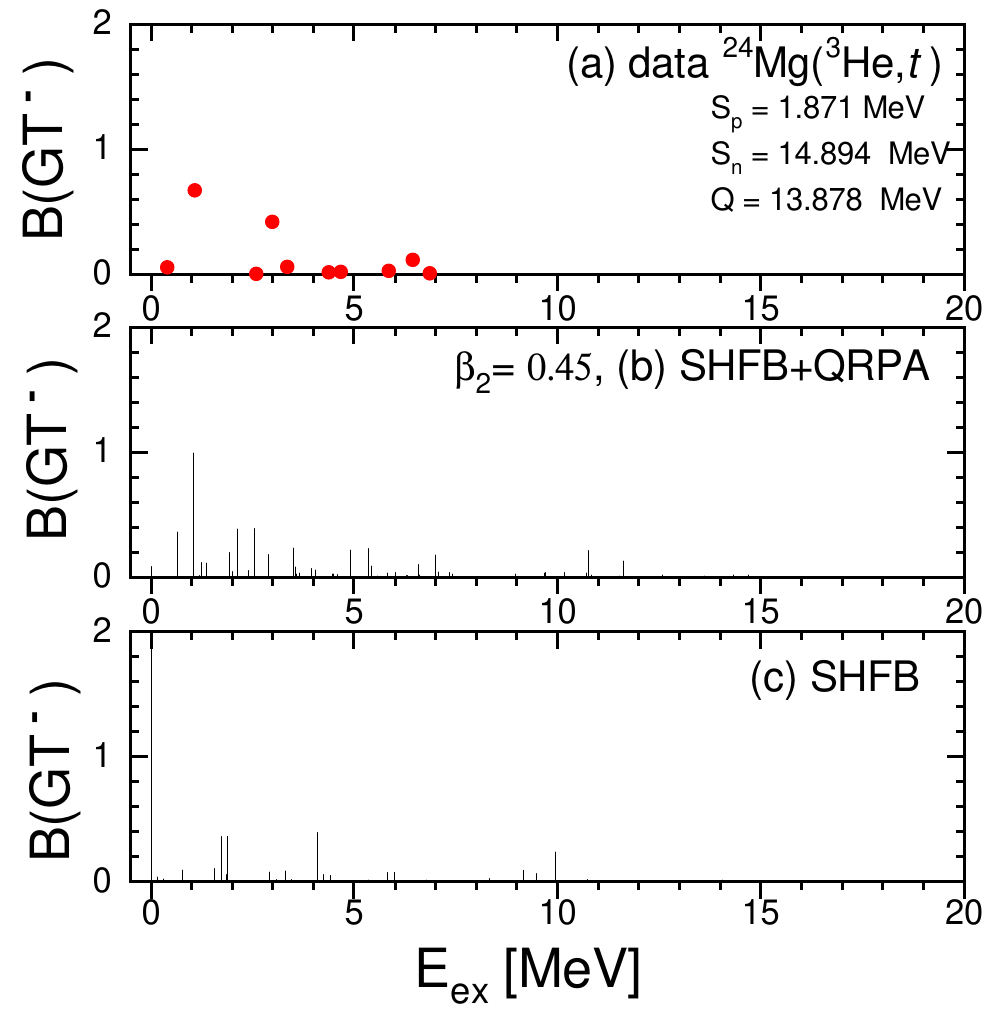}
\includegraphics[width=0.45\linewidth]{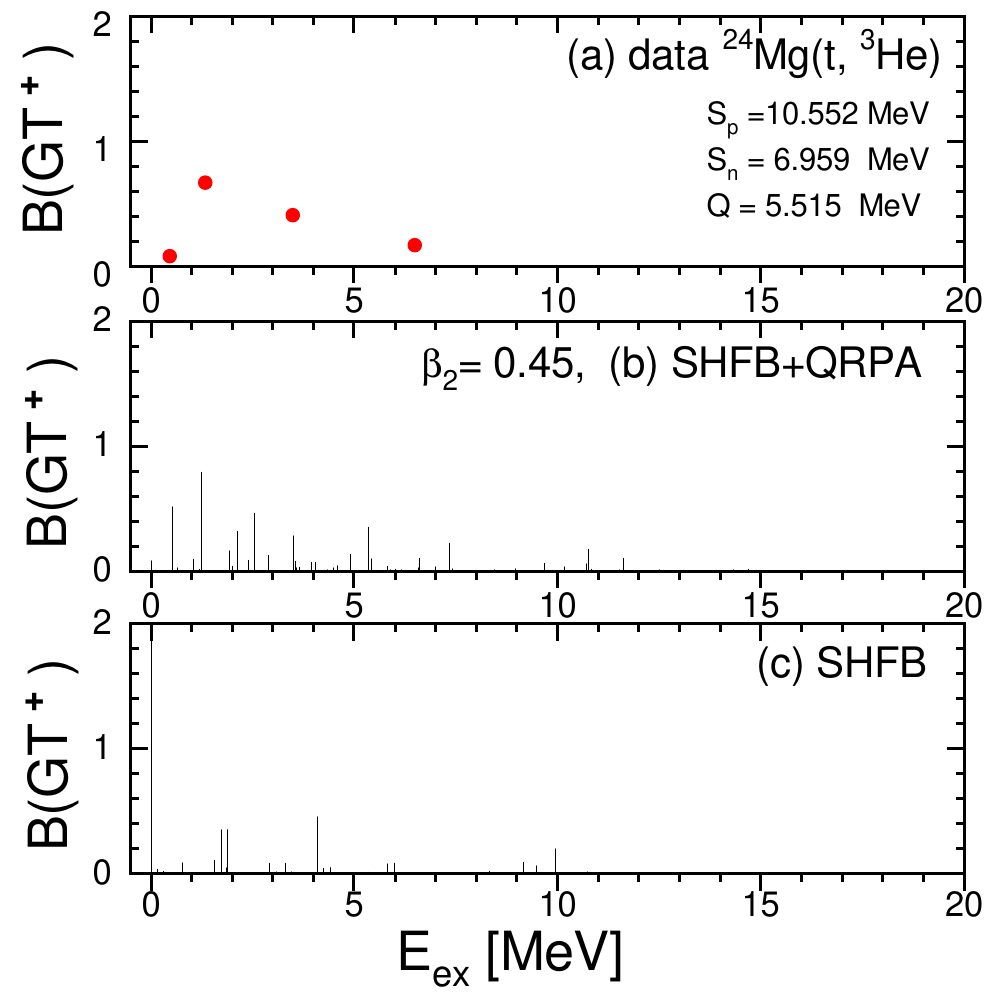}
\caption{(Color online) The B(GT$^{(\mp)}$) transition strength distributions  of $^{24}$Mg are displayed, respectively, in the left and right panels. Panels (c) correspond to the results of SHFB calculations without the residual interaction.}
\label{fig:24mg_gtmp}
\end{figure}
\begin{table}
\caption[bb]{Main configurations, type of configuration, and forward (backward) amplitudes for the first low-lying GT$^{(-)}$ state of $^{24}$Mg  at $E_{ex} \approx 0-1$ MeV in the left panel (b) of Fig. \ref{fig:24mg_gtmp}. {The GT transition {strength induced by the} $p-p$ interaction is shown to partially contribute to the low-lying GT peak.} \\ 
}
\setlength{\tabcolsep}{2.0 mm}
\begin{tabular}{cccc}\hline
                     &configuration (spherical limit)                                                & configuration type &  $X(Y)$        \\ \hline \hline
 SHFB+QRPA        &  $ \pi {1/2}_{3}^{+}(d3/2)$, $\nu {3/2}_{1}^{+}(d5/2)$  & $ p-h$     & 0.640 (0.022)   \\
                    &  $ \pi {1/2}_{3}^{+}(d3/2)$, $\nu {3/2}_{1}^{+}(d3/2)$  & $ p-p$     & 0.470 (0.019)   \\     
                    &  $ \pi {5/2}_{1}^{+}(d5/2)$, $\nu {3/2}_{1}^{+}(d5/2)$  & $ p-h$     & 0.450 (0.001)   \\
                    &  $ \pi {1/2}_{2}^{+}(d5/2)$, $\nu {3/2}_{1}^{+}(d3/2)$  & {$ h-p$}     & 0.260 (0.007)   \\\hline 
%   SHFB+rpa     &  $ \pi {1/2}_{3}^{+}(d3/2)$, $\nu {3/2}_{1}^{+}(d5/2)$  & $ p-h$     & 0.690(0.020)   \\
%  without TF   &  $ \pi {5/2}_{1}^{+}(d5/2)$, $\nu {3/2}_{1}^{+}(d5/2)$  & $ p-h$     & 0.410(0.002)   \\
%                    &  $ \pi {1/2}_{3}^{+}(d3/2)$, $\nu {3/2}_{1}^{+}(d3/2)$  & $ p-p$     & 0.370(0.008)   \\ 
%                    &  $ \pi {1/2}_{2}^{+}(d5/2)$, $\nu {3/2}_{1}^{+}(d3/2)$  & $ h-p$   & 0.250(0.009)   \\\hline   
    SHFB          &  $ \pi {1/2}_{3}^{+}(d3/2)$, $\nu {3/2}_{1}^{+}(d5/2)$  &     & 1.0 (0.0)   \\\hline \hline

 \end{tabular}
\label{tab:config1}
\end{table}
\begin{figure}
\includegraphics[width=0.45\linewidth]{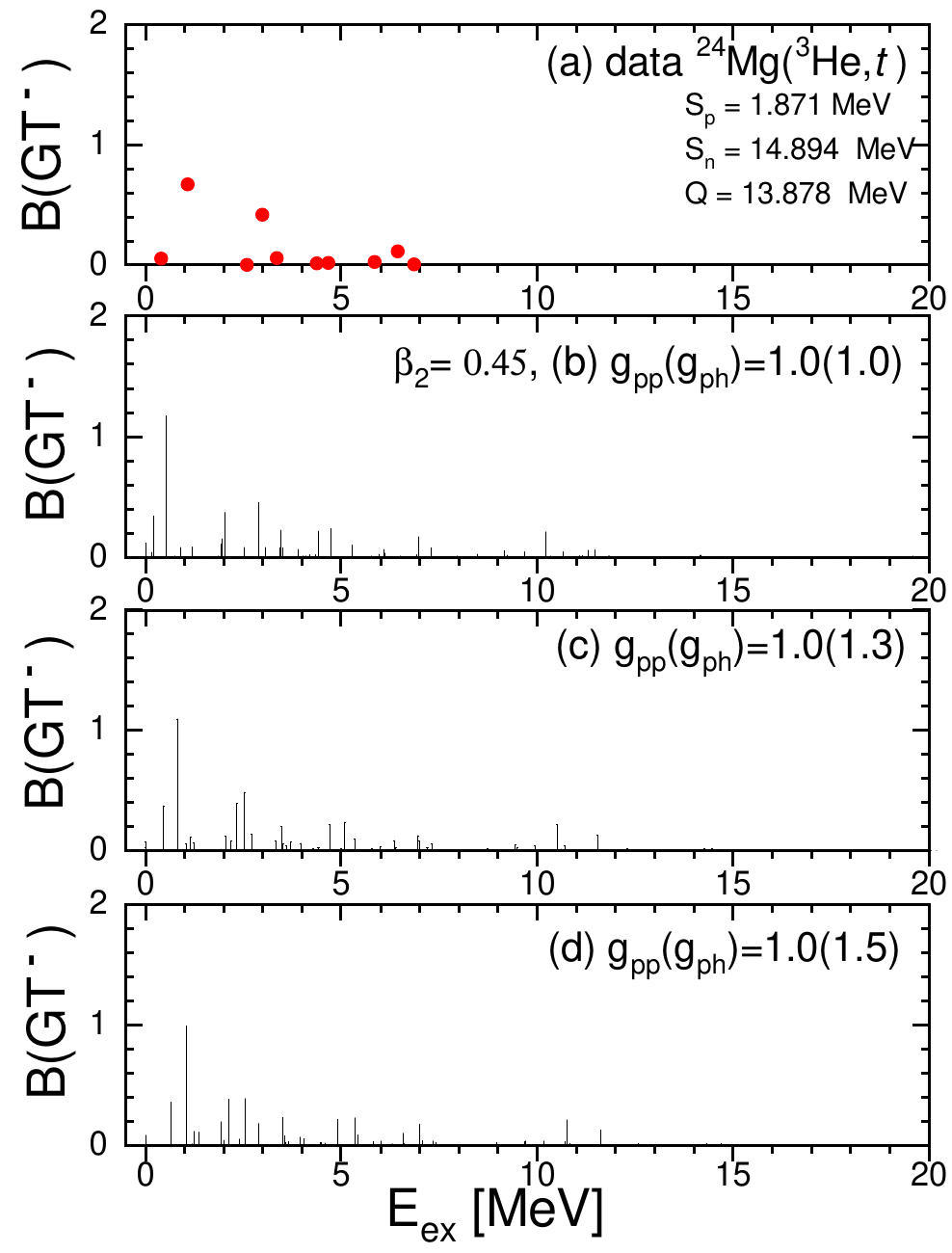}
\includegraphics[width=0.45\linewidth]{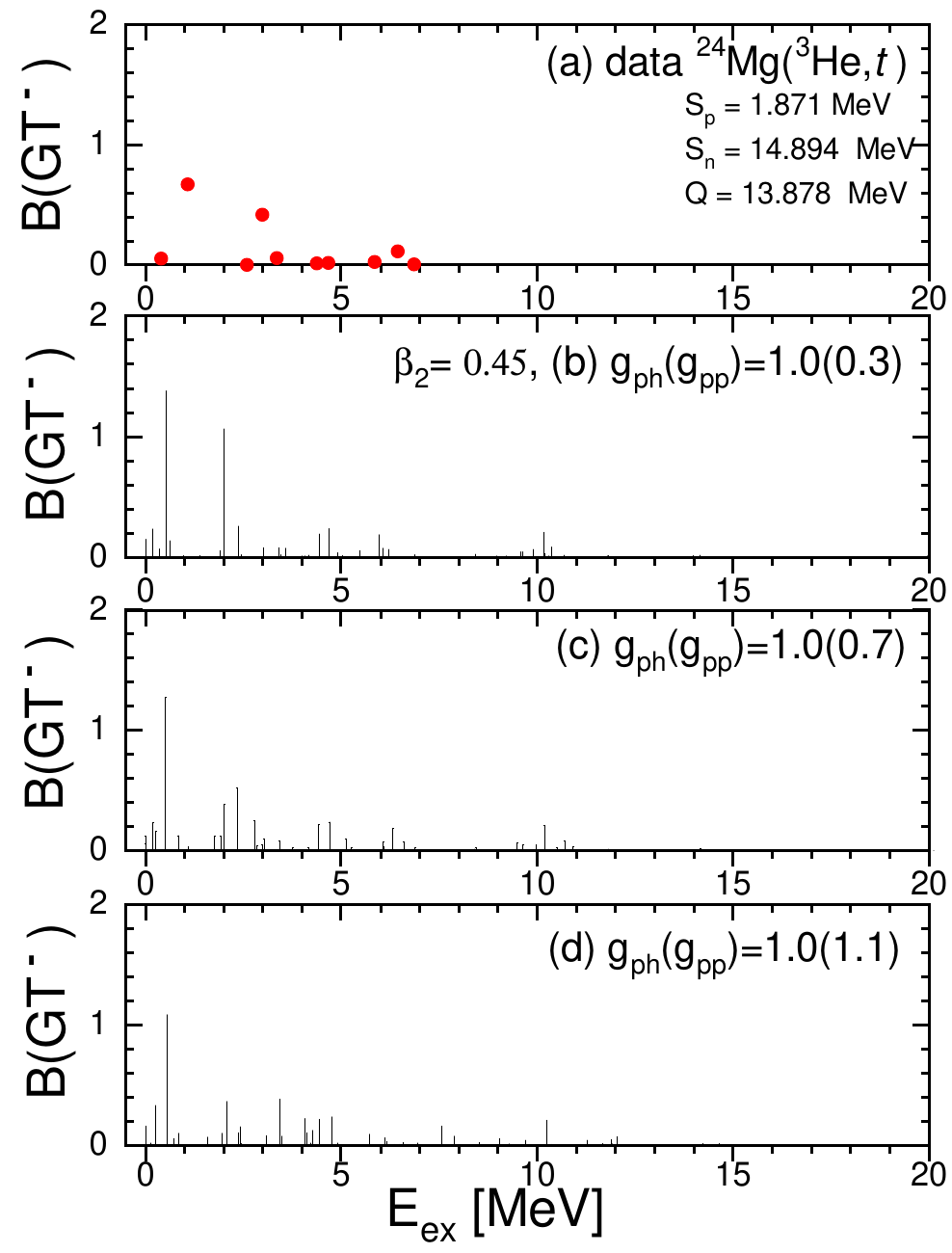}
\caption{(Color online) Effect of $p-h$ interaction on the GT$^{(-)}$ transition strength distributions of $^{24}$Mg. Results of (b), (c), and (d) in the left panel show the results where the normalization factor $g_{pp}$=1.0 is fixed in all the calculations, but the factor $g_{ph}$ is changed as 1.0, 1.3, and 1.5. Right panels show the results obtained by varying $g_{pp}$ as 0.3, 0.7 and 1.1 with a fixed  $g_{ph}$=1.0.}
\label{fig:24mg_gph}
\end{figure}
\begin{figure}
\includegraphics[width=0.55\linewidth]{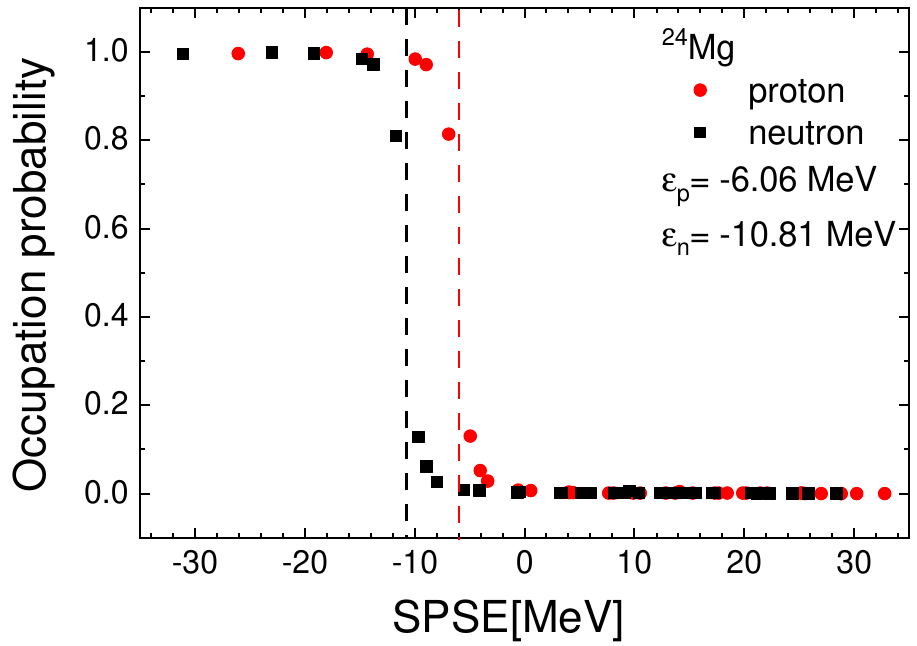}
\includegraphics[width=0.55\linewidth]{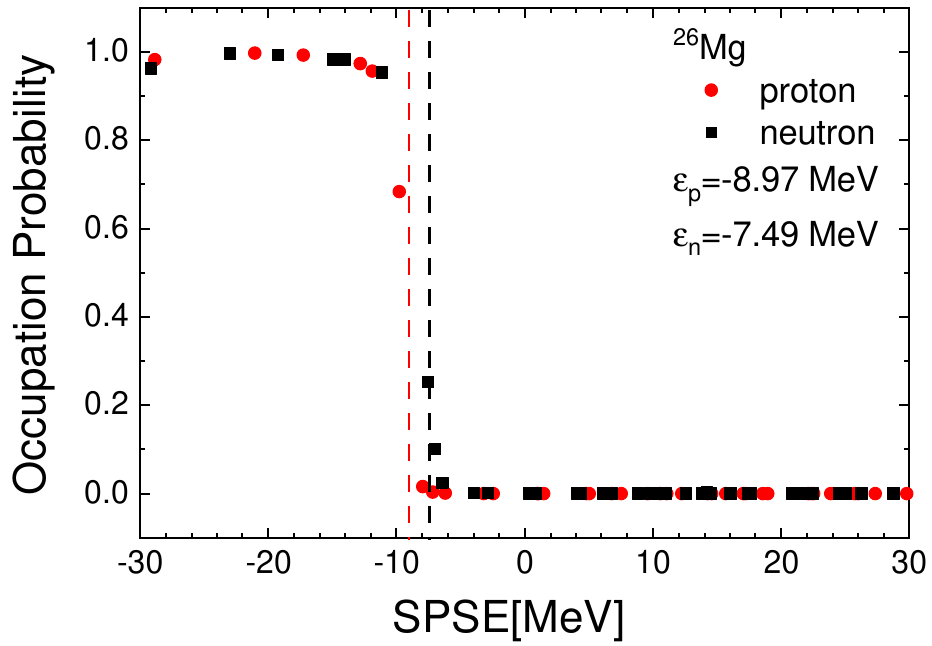}
\includegraphics[width=0.55\linewidth]{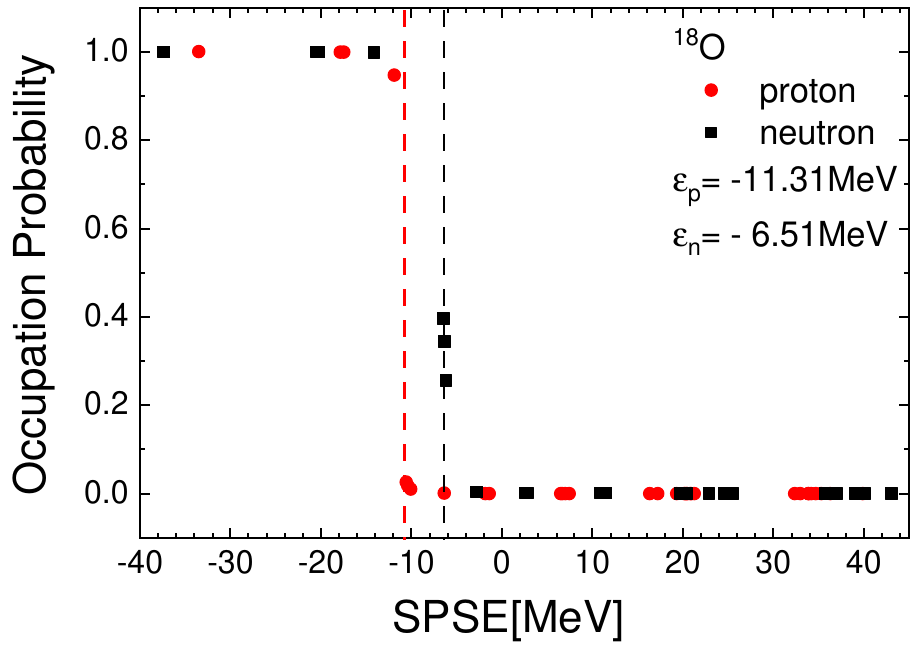}
\caption{(Color online) Occupation probabilities of protons and neutrons in $^{24}$Mg, $^{26}$Mg and $^{18}$O. The red (black) dashed line is the Fermi energy of protons (neutrons). Occupation probabilities of neutron SPSs around Fermi energy, $\nu d_{5/2} ({5/2}^+_1)$, $\nu s_{1/2} ({1/2}^+_3)$, and $\nu f_{7/2} ({1/2}^-_3)$ ($\nu d_{5/2} ({5/2}^+_1)$, $\nu d_{5/2} ({3/2}^+_1)$, and $\nu d_{5/2} ({1/2}^+_2)$) for $^{24,26}$Mg ($^{18}$O), are {smeared by the pairing interaction.}}
\label{fig:26mg_OP}
\end{figure}
\begin{figure}
\includegraphics[width=0.45\linewidth]{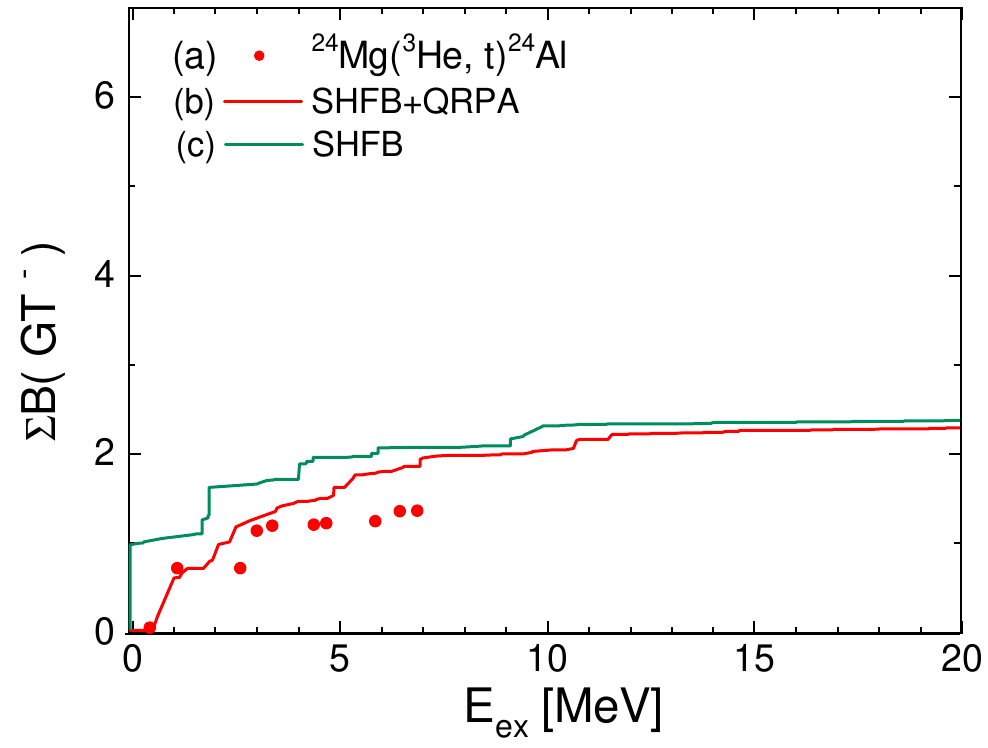}
\includegraphics[width=0.45\linewidth]{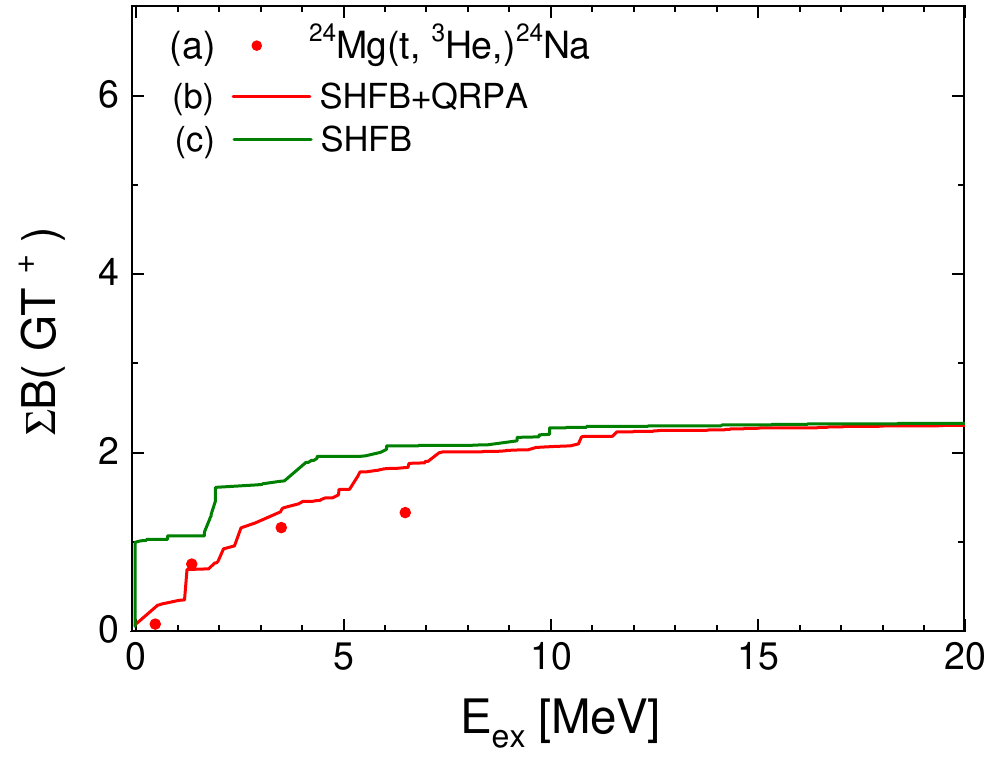}
\caption{(Color online) Cumulated  sum of the GT$^{(\mp)}$  transition strength distributions in $^{24}$Mg.  Here we use the quenching factor 0.684 for the GT operator (6), taken from Ref.~\cite{Kumar}.}
\label{fig:24mg_run}
\end{figure}

First, in Fig. \ref{fig:24mg_gtmp}, we present the results for $^{24}$Mg. The DQRPA explains GT$^{(-)}$ data as well as GT$^{(+)}$ data very well. Low energy GT peaks explicitly appear in the GT$^\mp$ strength distributions. It is interesting to see that both GT$^{(-)}$ and GT$^{(+)}$ strengths are concentrated in the low-energy region both {in experiment and theory}. 
The major  configuration of the low-energy peak of the GT$^{(-)}$ {strength} is the $\nu {3/2}_{1}^{+}(d5/2) \rightarrow \pi {1/2}_{3}^{+}(d3/2)$ $p-h$ excitation, while for the GT$^{(+)}$ peak, the 
$\pi {3/2}_{1}^{+}(d5/2) \rightarrow \nu {1/2}_{3}^{+}(d3/2)$  $p-h$ excitation is dominant.
This low-energy GT$^{(-)}$  strength may affect very much the $\beta$-decay and electron capture reactions in the relevant nucleosynthesis processes in the stellar burning.
A large {low-lying} GT$^{(-)}$ strength is also found in the doubly magic nucleus, 
 $^{48}$Ca \cite{Ha2023}, where about 15 \% of the sum rule  strength is shifted to the low energy GT peak because of the attractive TF interaction between $\pi f_{7/2}$ and $\nu f_{7/2}$ states in the framework of the DQRPA based on the WS potential.  Another interesting point to note is that the unperturbed configurations  dictate the feature of low-lying strengths  (see the panel (c) in Fig. \ref{fig:24mg_gtmp}), and  the residual QRPA interactions scatter the main peaks obtained by the unperturbed SHF mean field calculations.

If we assume that $^{24}$Mg is a closed shell nucleus with the sub-magic number $N=12$, the GT transition has to be dominated by the $p-h$ {configurations and residual} interaction. However some changes are induced by  the smearing of the Fermi surface  shown in Fig.~\ref{fig:26mg_OP},
associated  with the small $\Delta_n$ pairing gap in Table \ref{tab:pairing}. 
%the strong deformation is inducing some change: there is a slight , which is  and is due to the deformation  
Thus, the assumed sub-magic shell is weakened and  the effect of the $p-p$ interaction, {associated with}  $g_{pp}$, also appears explicitly, as shown in Table \ref{tab:config1}. In Fig.~\ref{fig:24mg_gph}, we examine the $p-h$ strength, or $g_{ph},$ dependence of the GT strength distribution. The increase of $g_{ph}$ slightly shifts the first peak to the higher energy region, and the change of $g_{pp}$ makes 
%does not \blu{produce} noticeable difference except 
a small change of the second peak strength, but the whole structure of the strength distribution is not changed. This is similar to the trend found in the case of double magic spherical nuclei like $^{48}$Ca, but the dominant $p-h$ interaction is a bit mixed with the $p-p$ interaction.

Fig.~\ref{fig:24mg_run} shows the cumulated  sum of the GT$^{(\mp)}$ distributions for $^{24}$Mg. We use the quenching factor {$q = 0.684$}, which is taken from a shell model study of nuclear $\beta^-$ decay half-lives in $fp$ and $fpg$ shell nuclei \cite{Kumar,Kumar2}. This is somewhat smaller than the commonly used quenching factor {$q_{univ} = 0.779$ \cite{Suzuki2023}.}     
{The adopted quenching factor reproduces well the gross feature of the cumulated sum, while the experimental data show further quenching than the calculated one. }

%%%%%%%%%%%%% 26Mg %%%%%%%%%%%%%%%%%%%%%%%%%%%%%%%%%%%%%%%%%%%%%%%%%%%%%%%%%%%%%%%%%%%%%%%%%%%%%
\subsection{$^{26}$Mg}
\begin{figure}
\includegraphics[width=0.45\linewidth]{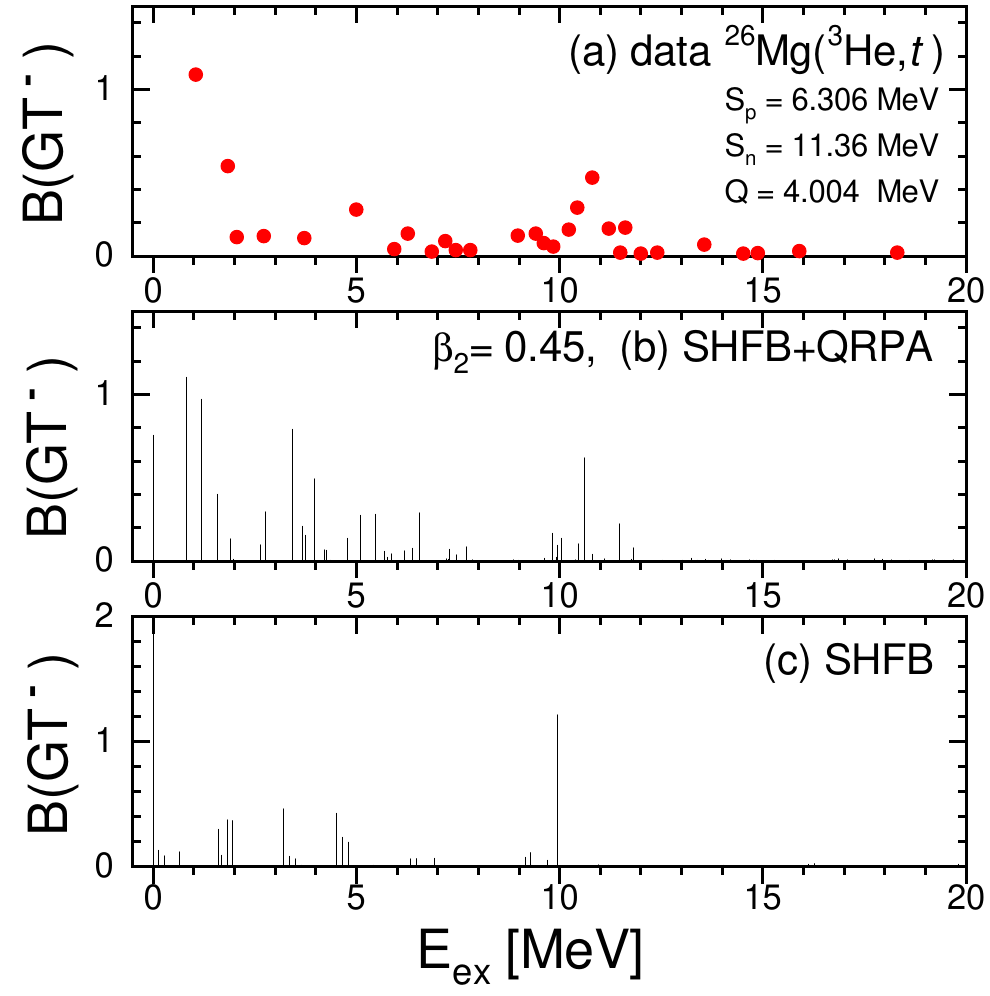}
\includegraphics[width=0.45\linewidth]{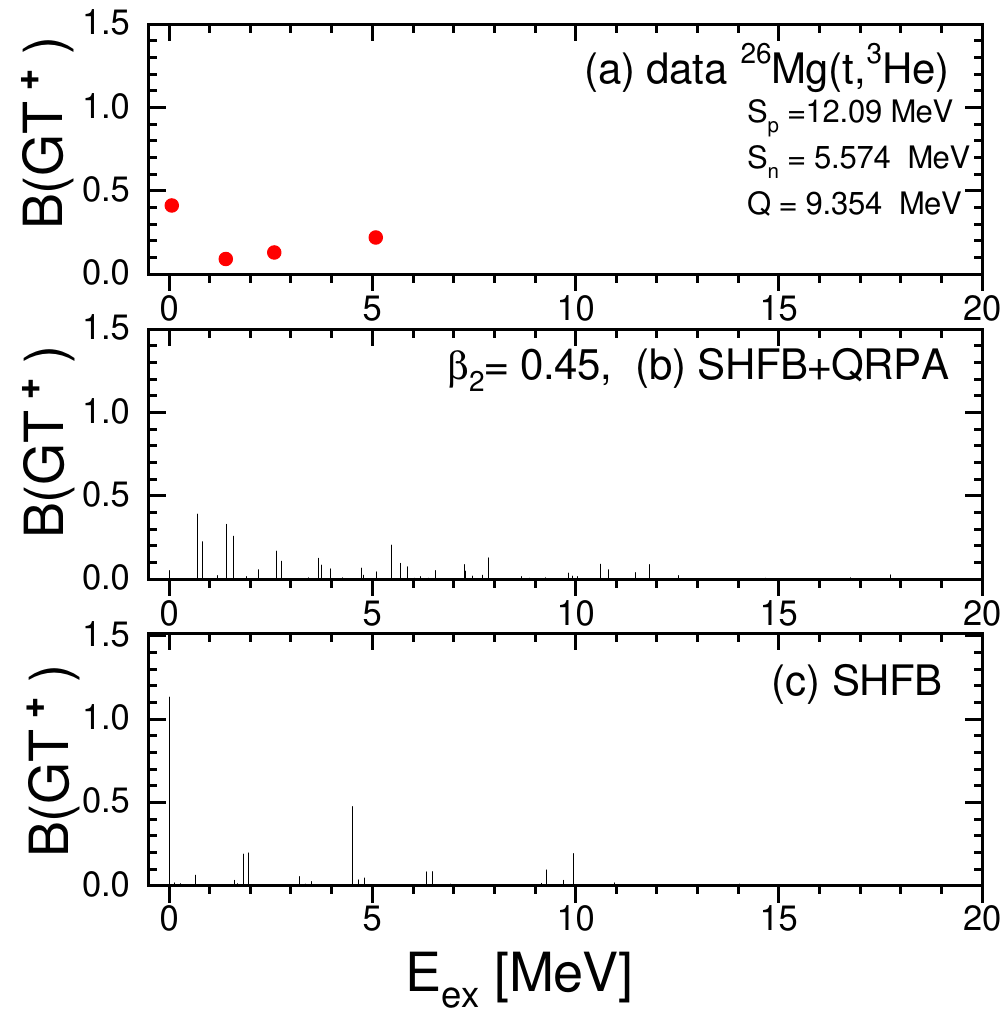}
\caption{(Color online) B(GT$^{\mp}$) %GT$^{(\mp)}$ 
 transition strength distribution of $^{26}$Mg. {Panels (c) results are from SHFB without residual interaction.}}
\label{fig:26mg_gtmp}
\end{figure}
\begin{table}
\caption[bb]{Main  configurations, type of configuration, and forward (backward) amplitude for the first low-lying GT$^{(-)}$ state of $^{26}$Mg at $E_{ex} \approx 1 $ MeV  in Fig. \ref{fig:26mg_gtmp}.\\ 
}
\setlength{\tabcolsep}{2.0 mm}
\begin{tabular}{cccc}\hline
                     &configuration (spherical limit)              & configuration type &  $X(Y)$        \\ \hline \hline
 SHFB+QRPA        &  $ \pi {3/2}_{1}^{+}(d5/2)$, $\nu {3/2}_{1}^{+}(d5/2)$  &  {$h-h$}      & 0.54 (0.16)   \\
                    &  $ \pi {1/2}_{3}^{+}(d3/2)$, $\nu {1/2}_{3}^{+}(d3/2)$  &  {$p-p$}    & 0.41 (0.14)   \\     
                    &  $ \pi {5/2}_{1}^{+}(d5/2)$, $\nu {5/2}_{1}^{+}(d5/2)$  & $ p-p$     & 0.28 (0.09)   \\\hline 
%  SHFB+rpa     &  $ \pi {3/2}_{1}^{+}(d5/2)$, $\nu {3/2}_{1}^{+}(d5/2)$  & $ p-h$     & 0.57(0.17)   \\
%without TF     &  $ \pi {1/2}_{3}^{+}(d3/2)$, $\nu {1/2}_{3}^{+}(d3/2)$  & $ p-h$     & 0.42(0.15)   \\     
%                    &  $ \pi {5/2}_{1}^{+}(d5/2)$, $\nu {5/2}_{1}^{+}(d5/2)$  & $ p-p$     & 0.22(0.11)   \\\hline  
    SHFB          &  $ \pi {1/2}_{3}^{+}(d3/2)$, $\nu {3/2}_{1}^{+}(d5/2)$  &     & 1.0 (0.0)   \\\hline \hline

 \end{tabular}
\label{tab:26config1}
\end{table}
\begin{table}
\caption[bb]{Main GT configurations, type of configuration, and forward (backward) amplitude for the high-lying GT$^{(-)}$ state of $^{26}$Mg at $E_{ex} \approx 10-11$ MeV in Fig. \ref{fig:26mg_gtmp}.\\ 
}
\setlength{\tabcolsep}{2.0 mm}
\begin{tabular}{cccc}\hline
                     &configuration (spherical limit)                                                & configuration type  &  $X(Y)$        \\ \hline \hline
 SHFB+QRPA        &  $ \pi {3/2}_{2}^{+}(d3/2)$, $\nu {5/2}_{1}^{+}(d5/2)$  & $ p-p$     & 0.49 (0.0035)   \\
                    &  $ \pi {1/2}_{3}^{-}(f7/2)$, $\nu {5/2}_{1}^{-}(f7/2)$  &  $p-p$      & 0.44 (0.0009)   \\     
                    &  $ \pi {3/2}_{1}^{-}(p3/2)$, $\nu {1/2}_{2}^{-}(p1/2)$  & $ h-h$     & 0.43 (0.0019)   \\\hline 
    SHFB          &  $ \pi {3/2}_{2}^{+}(d3/2)$, $\nu {5/2}_{1}^{+}(d5/2)$  &      & 1.0 (0.0)   \\\hline \hline

 \end{tabular}
\label{tab:26config2}
\end{table}
\begin{figure}
\includegraphics[width=0.45\linewidth]{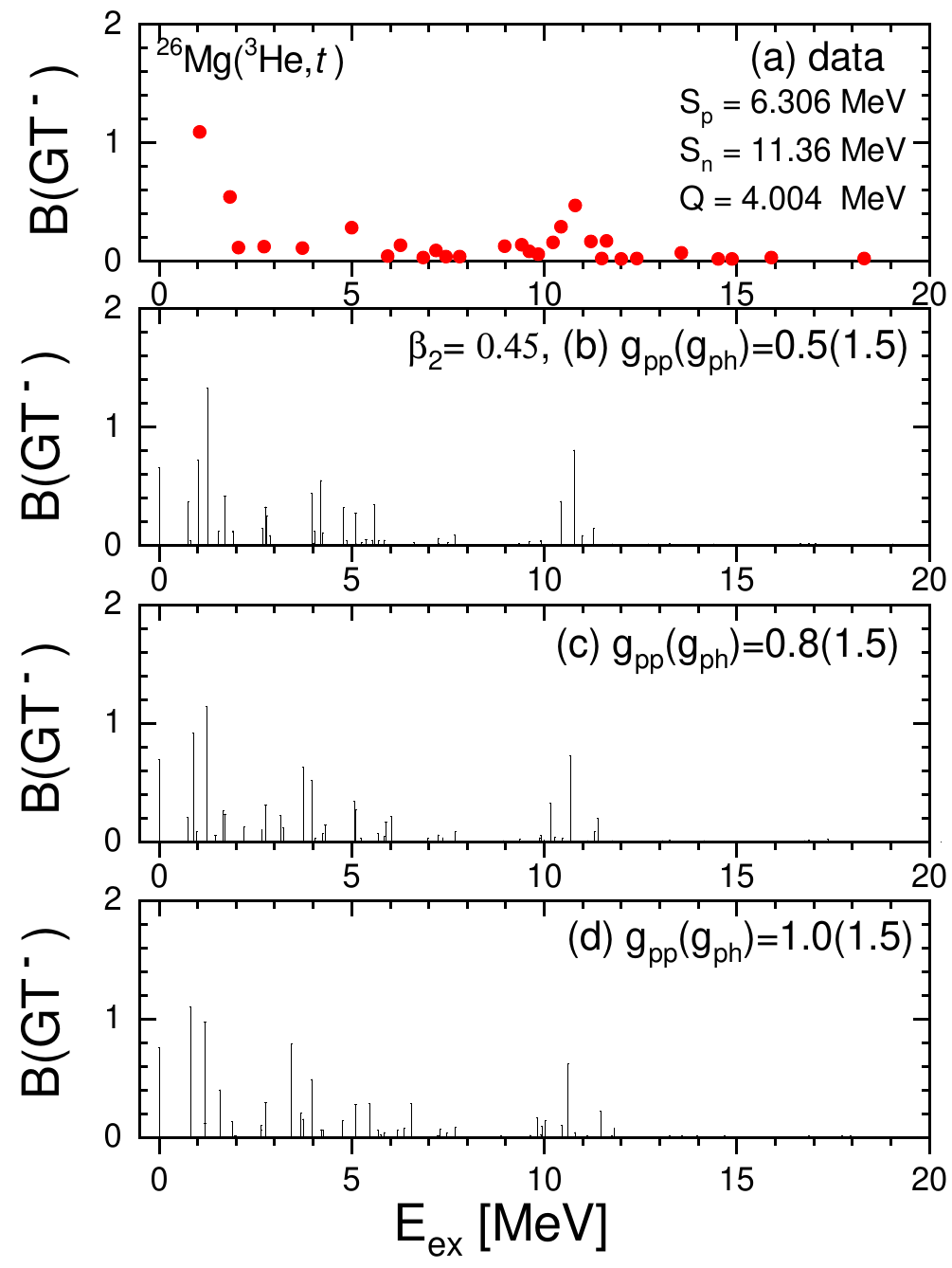}
\includegraphics[width=0.45\linewidth]{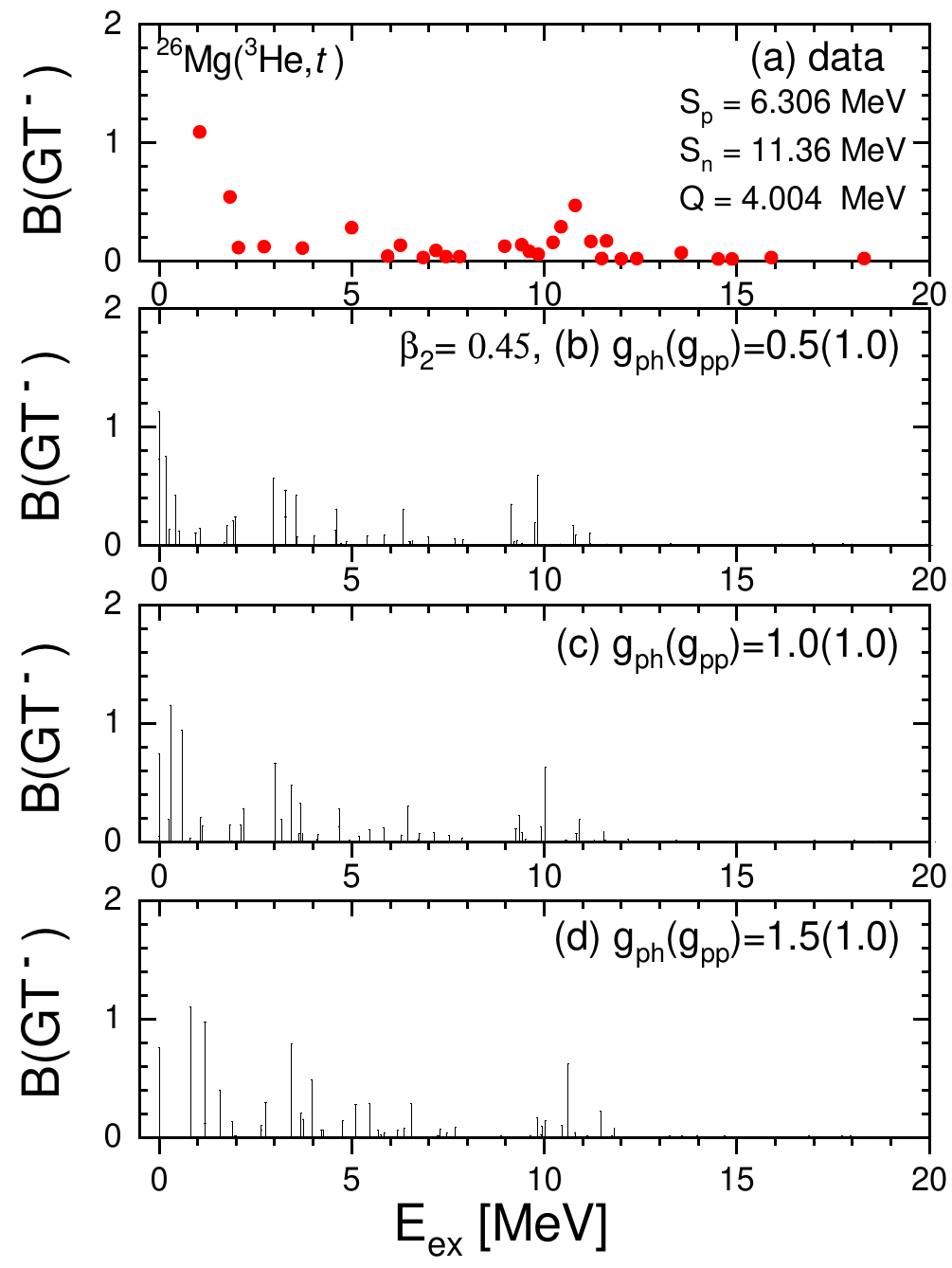}
\caption{(Color online) Effect of the $p-p$ interaction on the GT$^{(-)}$ transition strength distributions of $^{26}$Mg. Results of (b), (c), and (d) in the left panels correspond to the case in which the strength of the $p-h$ interaction, $g_{ph}$=1.5, is fixed but the strength of the $p-p$ interaction $g_{pp}$ is changed as 0.5, 0.8, and 1.0. The right panels shows the results changing $g_{ph}$ as 0.5, 1.0, and 1.5, by keeping  $g_{pp}$=1.0.}
\label{fig:26mg_gpp}
\end{figure}

\begin{figure}
\includegraphics[width=0.45\linewidth]{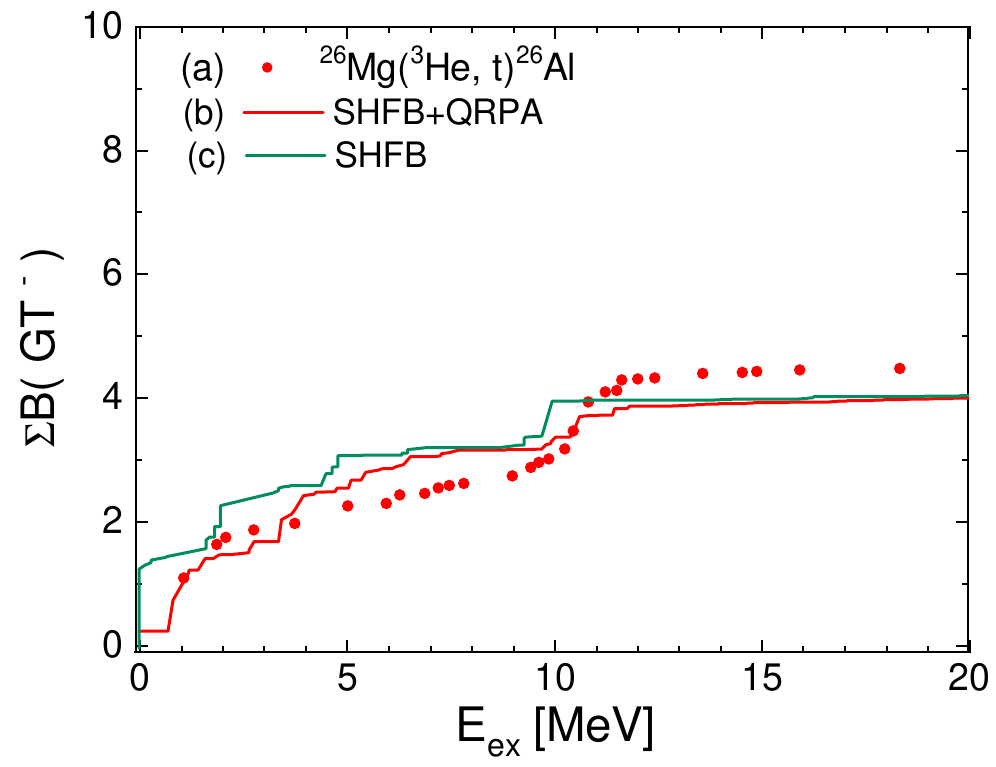}
\includegraphics[width=0.45\linewidth]{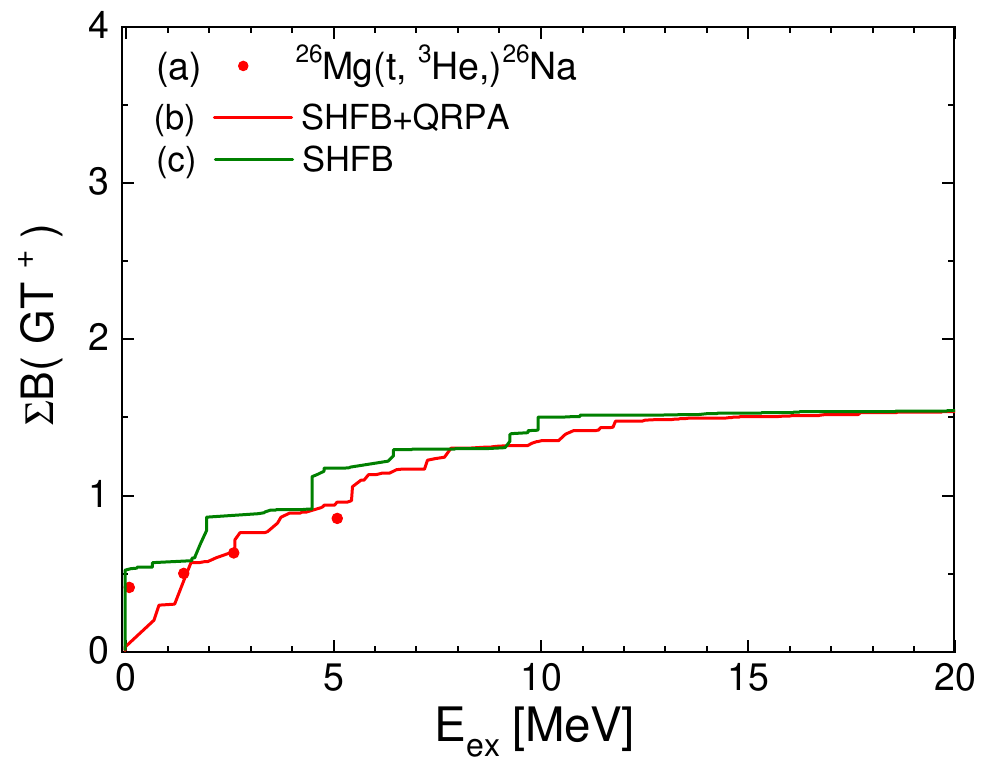}
\caption{(Color online) Cumulated  sum of GT$^{(\mp)}$ transition strength distributions in $^{26}$Mg.}
\label{fig:26mg_run}
\end{figure}

\vskip1cm
Hereafter, we discuss the results for $^{26}$Mg, where one can find a low-lying GT peak as well as the typical GTGR peak around 12 MeV as shown in Fig. \ref{fig:26mg_gtmp}(a). This strength distribution is similar to the case of $^{42}$Ca \cite{Ha2023}, {\it i.e.}, a nucleus with $N = Z+2$.

Figure \ref{fig:26mg_gtmp} (b) and (c) show numerical results for the  B(GT$^{(\mp)}$) strength distribution for $^{26}$Mg: {we remind its large deformation, associated with} $\beta_2 =0.45$. The experimental data are described well by the present calculations. The main configurations for the first peak at $E_{ex} \approx 1$ MeV  in Fig.~\ref{fig:26mg_gtmp} (b) and (c) are presented in Table \ref{tab:26config1}. 
Compared with the $^{24}$Mg case, there are 2 extra neutrons above the semi-magic shell $N=12$, which will create more 
$p-p$ type configurations than in $^{24}$Mg.
However, because of the the smearing due to the pairing interaction as shown in Fig.~\ref{fig:26mg_OP}, the $p-h$ interaction channels also play  an important role as well as  the $p-p$ one.
This {can be seen} in Fig. \ref{fig:26mg_gpp}, in which the effect of the variation of $g_{pp}$ is similar  to  that induced by the change of $g_{ph}$: the lower energy strengths are changed appreciably, but the higher-energy peak is not so much changed by the increase of $g_{pp}$ in the left panels of  Fig. \ref{fig:26mg_gpp}.  A similar effect is also found in the right panels where the value $g_{ph}$ is increased.

%%%%% 3:35 Feb. 15

Now, we will discuss detailed features of  the GT transitions in Tables \ref{tab:26config1} and \ref{tab:26config2}. The first peak comes from two transitions between $ \pi {3/2}_{1}^{+}(d5/2)$ and $\nu {3/2}_{1}^{+}(d5/2)$, and $ \pi {1/2}_{3}^{+}(d3/2)$ and $\nu {1/2}_{3}^{+}(d3/2)$. We note that the low-lying GT transitions stem from the same $j$ states of protons and neutrons in the spherical limit, {\it i.e.},  transitions within $d5/2$ states as well as within $d3/2$ spherical states.
 
The main configurations of the high-lying GT state at $E_{ex}\approx 10-11$ MeV  in Fig.~\ref{fig:26mg_gtmp} (b)-(d)  are presented in Table \ref{tab:26config2}, which shows the transition between different states {\it i.e}, $ \pi {3/2}_{2}^{+}(d3/2)$ and $\nu {5/2}_{1}^{+}(d5/2)$ and $ \pi {3/2}_{1}^{-}(p3/2)$ and $\nu {1/2}_{2}^{-}(p1/2)$.  
It is interesting to notice that the negative parity  configuration $\nu {1/2}_{2}^{-}(p1/2) \rightarrow \pi {3/2}_{1}^{-}(p3/2)$ has also a large RPA amplitude to enhance the GT strength.
{The cumulated sum is shown in Fig. 8.  The gross feature  of the sum is reproduced by our calculations of SHFB+QRPA with the quenching factor {$q$=0.684.}}

%%%%%%%%%%%%%%%%%%%% 18O %%%%%%%%%%%%%%%%%%%%%%%%%%%%%%%%%%%%%%%%%%%%%%%
\subsection{$^{18}$O}
\begin{figure}
\includegraphics[width=0.45\linewidth]{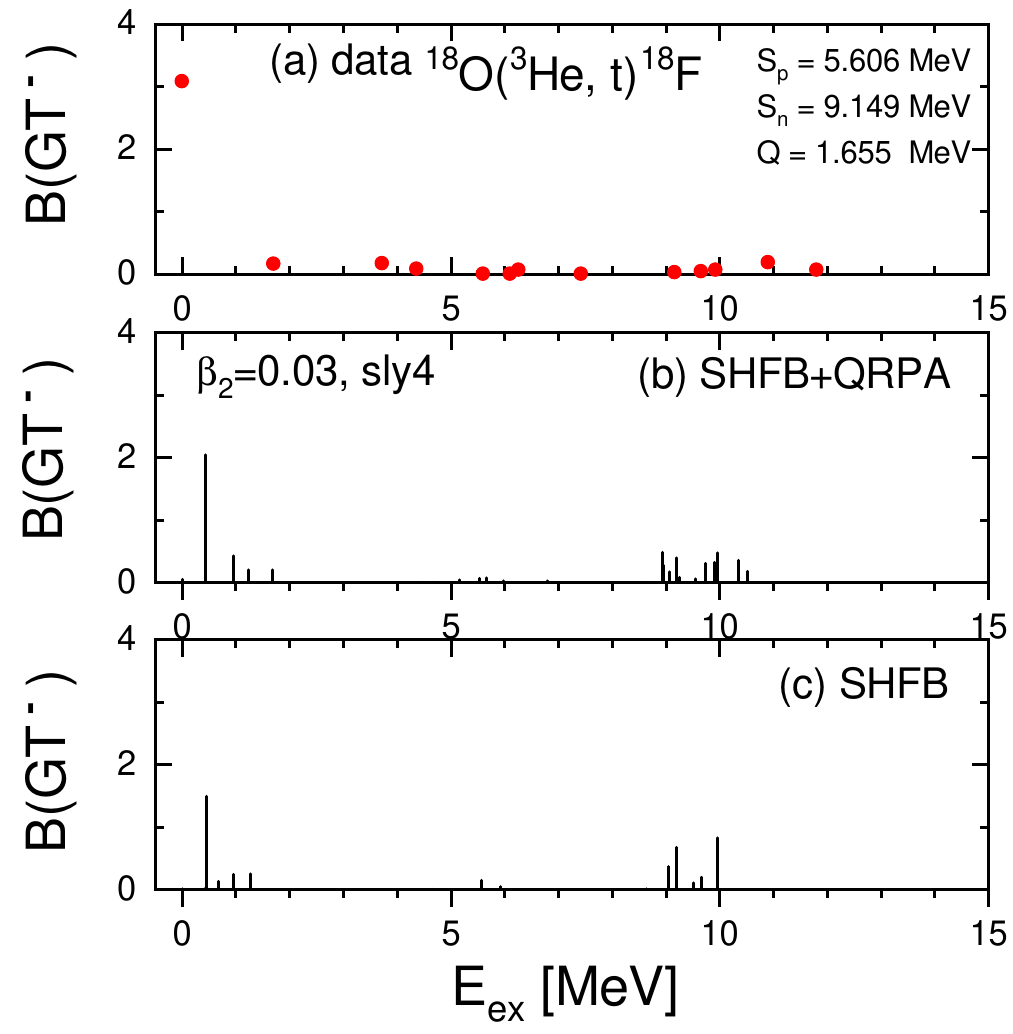}
\caption{(Color online)  B(GT$^{(-)}$)  transition strength distribution of $^{18}$O. Results of (b) and (c) are obtained by SHFB + {DQRPA} and by SHFB, respectively.}
\label{fig:18O_gtmp}
\end{figure}

\begin{table}
\caption[bb] {
{Main  configurations, type of configuration, and forward (backward) amplitude for} the low-lying GT state of $^{18}$O at $E_{ex} \approx 0.5$ MeV in the right panel of Fig. \ref{fig:18O_gtmp}.\\
}
\setlength{\tabcolsep}{2.0 mm}
\begin{tabular}{cccc}
                     &configuration (spherical)                                                & {configuration type} &  $X(Y)$        \\ \hline \hline
  SHFB+QRPA     & $ \pi {1/2}_{2}^{+}(d5/2)$, $\nu {1/2}_{2}^{+}(d5/2)$  &  $ p-p$      &0.57 (0.08)  \\
                      & $ \pi {3/2}_{1}^{+}(d5/2)$, $\nu {3/2}_{1}^{+}(d5/2)$  &  $ p-p$     &0.40 (0.07)  \\    
                      & $ \pi {1/2}_{2}^{-}(p1/2)$, $\nu {1/2}_{2}^{-}(p1/2)$  &  $ h-h$     & 0.29 (0.04)   \\
                      &  $ \pi {5/2}_{1}^{+}(d5/2)$, $\nu {5/2}_{1}^{+}(d5/2)$  & $ p-p$     & 0.22 (0.06)   \\\hline                 
    SHFB          &  $ \pi {1/2}_{2}^{+}(d5/2)$, $\nu {1/2}_{2}^{+}(d5/2)$  &               &  1.0 (0.0)     \\\hline \hline 
 \end{tabular}
\label{tab:config3}
\end{table}
\begin{figure}
\includegraphics[width=0.45\linewidth]{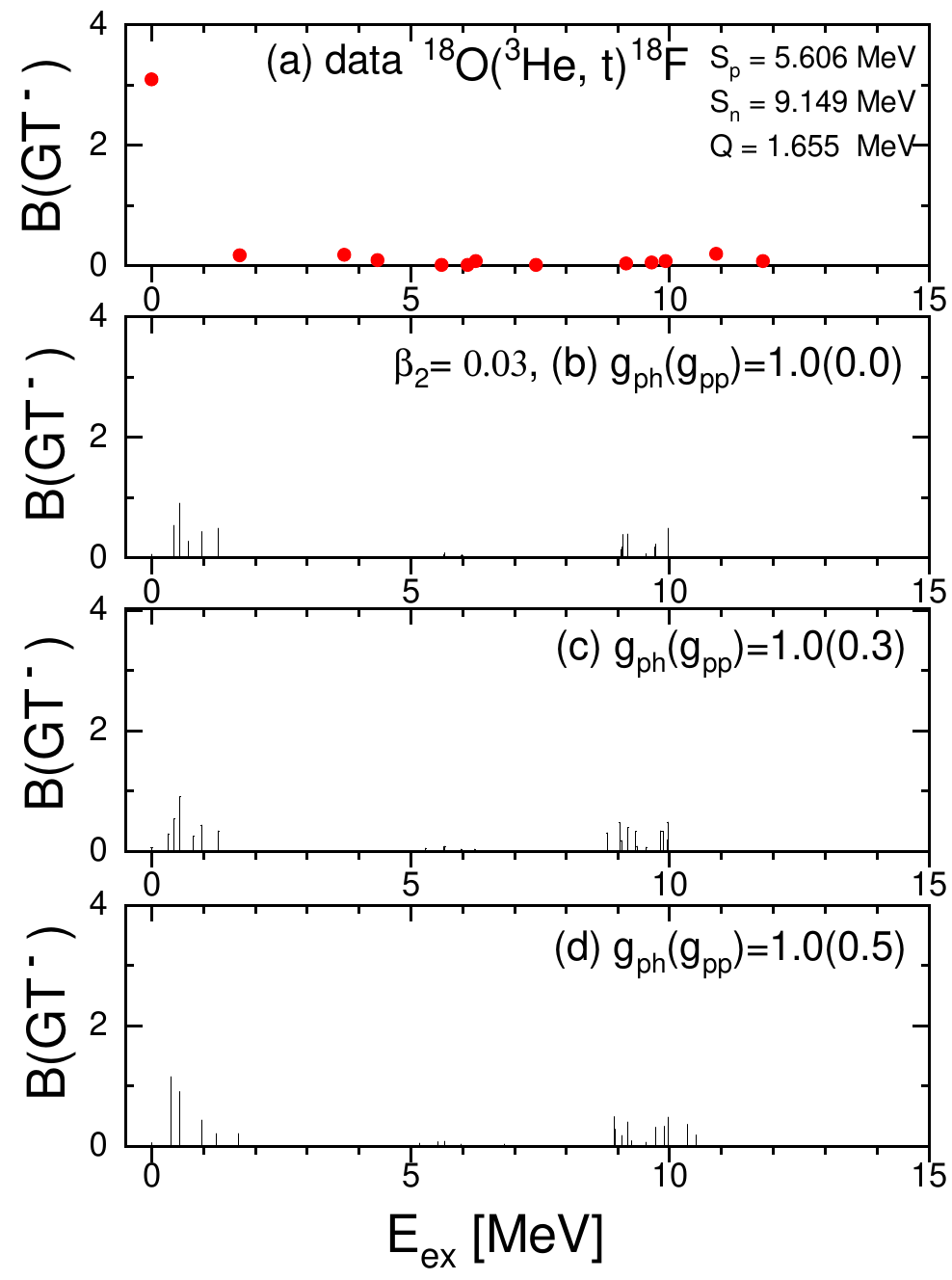}
\includegraphics[width=0.45\linewidth]{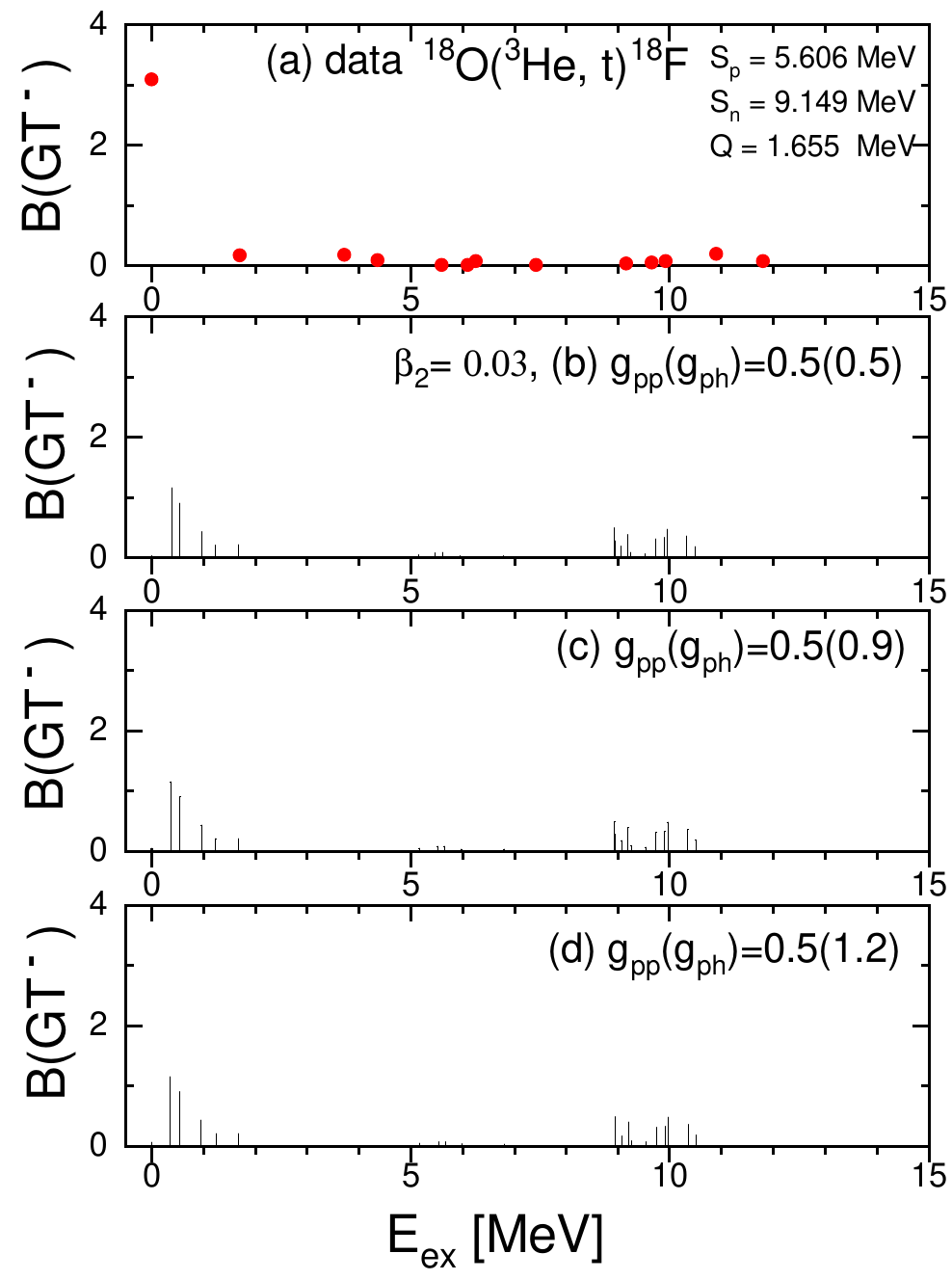}
\caption{(Color online) Effects of $p-p$ and $p-h$ interaction on the GT$^{(-)}$ transition strength distributions of $^{18}$O. Results of (b), (c), and (d) in the left panel show the results obtained by fixing $g_{ph}$=1.0, with $g_{pp} =$ 0.0, 0.3, and 0.5. Those in the right panel are for $g_{pp} = 0.5$, with $g_{ph}$ = 0.5, 0.9, and 1.2.}
\label{fig:18Omg_gpp}
\end{figure}

Finally, we discuss the results for $^{18}$O, which has a negligible intrinsic deformation in the DHF calculation and no rotational band structure was found experimentally, while the vibrational states are found \cite{Shukla}. Therefore we consider the SHFB solution associated with $\beta_2 = 0.03$ in the following. We also expect that the GT strength distribution of $^{18}$O has low-lying GT peaks similar to those in $^{26}$Mg, because of the two extra neutrons on top of the magic number $N=8$. Figure \ref{fig:18O_gtmp} shows  results of the B(GT$^{(-)}$) strength distribution for $^{18}$O.
The results show low-lying peaks and also some GT strength  around 10 MeV. The results are compatible with the experimental data.

The main configurations for  the first peak at $E_{ex} \approx 0.5$ MeV  in Fig.~\ref{fig:18O_gtmp}(b) are presented in Table \ref{tab:config3}. This  peak comes, largely, from the configuration $\nu {1/2}_{2}^{+}(d5/2)  
\rightarrow \pi {1/2}_{2}^{+}(d5/2)$, under the influence of the attractive TF interaction. 
It turns out that a low-lying super-GT peak is also found in $^{18}$O similarly to other $N=Z+2$ nuclei such as  $^{42}$Ca \cite{Fujita2015} and $^{46}$Ti \cite{Adachi2006}. Since the GT results of the $N=Z+2$ nuclei are sensitive to the $g_{pp}$ strength we also provide the sensitivity tests to $g_{pp}$ and $g_{ph}$ in Fig. \ref{fig:18Omg_gpp}. They show clearly the $g_{pp}$ dependence in the low-lying GT peaks in the left panels, while the
$g_{ph}$ dependence  is almost invisible in the right panels. Another interesting point is that $^{18}$O does not show GTGR. It may come from the fact that the GTGR {strength} is fully shifted to low-lying GT states by a strong $p-p$ residual interaction, since the spin-orbit force is weaker than the case of  $^{26}$Mg. However, there are some differences between data and theory in the $2 < E_{ex} < 8 $ MeV region. We note that the $( {^3}He, t)$ experiments may have strong background compared to those by $(p,n)$ reaction, and consequently, more careful analysis might be necessary to confirm the strength in the $2 < E_{ex} < 8 $ MeV region.

\section{Summary and conclusion}

We investigated the GT strength distributions of largely deformed s-d shell nuclei, $^{24}$Mg, $^{26}$Mg, and {of an almost spherical} nucleus, $^{18}$O,  in the framework of Deformed QRPA (DQRPA).  
 The present calculation adopts  the SHF mean field with the residual interaction derived the $G$-matrix calculation. 
 We found that {major part} of the experimental data {are better explained  than in} the previous calculations based on the Woods-Saxon potential \cite{Ha2016}. {We found that  $^{24}$Mg and $^{26}$Mg are strongly deformed, and} a sub-magic shell at $N=12$   is found and plays an important role to describe the GT strength distributions of these nuclei. {The cumulated sums of GT$_{\pm}$ strengths of $^{24}$Mg, $^{26}$Mg are studied in the low-energy region where the experimental data are available. Our SHFB+QRPA results give a good account of the sum rule values with a quenching factor $q=0.684$ for the GT operator.} 

{We also confirmed similarities of the GT strength distributions of $N=Z$ nuclei to those of  $N=Z+2$ nuclei.} It implies that the low-lying GT states in $N=Z$ $(N=Z+2)$ nuclei are enhanced mainly by the {$p-h$ ($p-p$) interaction},  and relatively small GT strengths appear in the higher energy region in the case of the nuclei studied in this article. We argued that the strong deformation {makes} {it so that the relevant residual interaction 
is a} combination of $p-p$ and $p-h$ interactions, as shown in $^{24}$Mg and $^{26}$Mg, due to the smearing of Fermi surface by  the pairing correlations. {On the other hand,} this phenomenon is not significant in the case of $^{18}$O nucleus, {since it is almost spherical.}

Finally, we note that the present DQRPA formalism is constructed to include the $T=0$ pairing on top of the $T=1$ pairing correlations, although the $T=0$ pairing is not considered in this work. In particular, a group of nuclei {approximately} with neutron number $60< N <70$ near the proton drip-line, such as $^{132}$Dy and $^{132}$Nd, have been reported to {be good candidates for displaying} both $T=0$ and $T=1$ pairing correlations, and consequently may have low-lying excitations associated with spin mixing, due to both isospin {pairing} correlations \cite{Alex2011}. Furthermore, recently $\beta$-delayed spectroscopy from nuclei near the proton drip-line, for example, $^{56}$Zn, $^{48}$Fe and $^{60,62}$Ge, are available \cite{Sonj2023}. We leave the application {of our formalism} to the nuclei near proton drip-line as a future work.

\section*{Acknowledgement}
This work was supported by the National Research Foundation of Korea (Grant Nos. NRF-2018R1D1A1B05048026).  The work of MKC is supported by the National Research Foundation of Korea (Grant Nos. NRF-2021R1A6A1A03043957 and NRF-2020R1A2C3006177). 

\end{document}